\documentclass[pdflatex,sn-nature,oneside]{sn-jnl}

\usepackage{graphicx}
\usepackage{multirow}
\usepackage{amsmath,amssymb,amsfonts}
\usepackage{amsthm}
\usepackage{mathrsfs}
\usepackage[title]{appendix}
\usepackage{xcolor}
\usepackage{textcomp}
\usepackage{manyfoot}
\usepackage{booktabs}
\usepackage{algorithm}
\usepackage{algorithmicx}
\usepackage{algpseudocode}
\usepackage{listings}
\usepackage[labelformat=empty]{caption}
\usepackage{braket}
\usepackage{geometry}
\usepackage{fancyhdr}
\begin{document}
\vspace{-3cm}
\title[Article Title]{Symmetry-protected electronic metastability in an optically driven cuprate ladder}


\author*[1]{\fnm{Hari} \sur{Padma}}\email{hpadmanabhan@g.harvard.edu}
\author[1]{\fnm{Filippo} \sur{Glerean}}
\author[1,2]{\fnm{Sophia} \sur{F. R. TenHuisen}}
\author[3]{\fnm{Zecheng} \sur{Shen}}
\author[4]{\fnm{Haoxin} \sur{Wang}}
\author[3]{\fnm{Luogen} \sur{Xu}}
\author[5]{\fnm{Joshua} \sur{D. Elliott}}
\author[6]{\fnm{Christopher} \sur{C. Homes}}
\author[7]{\fnm{Elizabeth} \sur{Skoropata}}
\author[7]{\fnm{Hiroki} \sur{Ueda}}
\author[7]{\fnm{Biaolong} \sur{Liu}}
\author[7]{\fnm{Eugenio} \sur{Paris}}
\author[7]{\fnm{Arnau} \sur{Romaguera}}
\author[8,9]{\fnm{Byungjune} \sur{Lee}}
\author[10]{\fnm{Wei} \sur{He}}
\author[11,12]{\fnm{Yu} \sur{Wang}}
\author[11,12]{\fnm{Seng Huat} \sur{Lee}}
\author[13]{\fnm{Hyeongi} \sur{Choi}}
\author[13]{\fnm{Sang-Youn} \sur{Park}}
\author[11,12]{\fnm{Zhiqiang} \sur{Mao}}
\author[14]{\fnm{Matteo} \sur{Calandra}}
\author[13]{\fnm{Hoyoung} \sur{Jang}}
\author[7]{\fnm{Elia} \sur{Razzoli}}
\author[10]{\fnm{Mark P. M.} \sur{Dean}}
\author*[3]{\fnm{Yao} \sur{Wang}}\email{yao.wang@emory.edu}
\author*[1]{\fnm{Matteo} \sur{Mitrano}}\email{mmitrano@g.harvard.edu}

\affil[1]{\orgdiv{Department of Physics}, \orgname{Harvard University}, \orgaddress{\city{Cambridge}, \state{MA}, \country{USA}}}

\affil[2]{\orgdiv{Department of Applied Physics}, \orgname{Harvard University}, \orgaddress{\city{Cambridge}, \state{MA}, \country{USA}}}

\affil[3]{\orgdiv{Department of Chemistry}, \orgname{Emory University}, \orgaddress{\city{Atlanta}, \state{GA}, \country{USA}}}

\affil[4]{\orgdiv{Department of Physics}, \orgname{The Chinese University of Hong Kong}, \orgaddress{\city{Hong Kong}}}

\affil[5]{\orgname{Diamond Light Source}, \orgaddress{\city{Didcot}, \state{Oxfordshire}, \country{UK}}}

\affil[6]{\orgdiv{National Synchrotron Light Source II}, \orgname{Brookhaven National Laboratory}, \orgaddress{\city{Upton}, \state{NY}, \country{USA}}}

\affil[7]{\orgname{PSI Center for Photon Science}, \orgname{Paul Scherrer Institute}, \orgaddress{\city{Villigen}, \country{Switzerland}}}

\affil[8]{\orgdiv{Department of Physics}, \orgname{Pohang University of Science and Technology}, \orgaddress{\city{Pohang}, \country{Korea}}}

\affil[9]{\orgdiv{Max Planck POSTECH/Korea Research Initiative}, \orgname{Center for Complex Phase Materials}, \orgaddress{\city{Pohang}, \country{Korea}}}

\affil[10]{\orgdiv{Condensed Matter Physics and Materials Science Department}, \orgname{Brookhaven National Laboratory}, \orgaddress{\city{Upton}, \state{NY}, \country{USA}}}

\affil[11]{\orgname{Department of Physics}, \orgname{Pennsylvania State University}, \orgaddress{\city{University Park}, \state{PA}, \country{USA}}}

\affil[12]{\orgname{2D Crystal Consortium}, \orgname{Materials Research Institute}, \orgname{Pennsylvania State University}, \orgaddress{\city{University Park}, \state{PA}, \country{USA}}}

\affil[13]{\orgname{Pohang Accelerator Laboratory (PAL)}, \orgname{Pohang University of Science and Technology}, \orgaddress{\city{Pohang}, \country{South Korea}}}

\affil[14]{\orgname{Department of Physics}, \orgname{University of Trento}, \orgaddress{\city{Povo}, \country{Italy}}}

\baselineskip24pt %

\maketitle

\textbf{Optically excited quantum materials exhibit nonequilibrium states with remarkable emergent properties, but these phenomena are usually transient, decaying on picosecond timescales and limiting practical applications. Advancing the design and control of nonequilibrium phases requires the development of targeted strategies to achieve long-lived, metastable phases. Here, we report the discovery of symmetry-protected electronic metastability in the model cuprate ladder Sr$_{14}$Cu$_{24}$O$_{41}$. Using femtosecond resonant x-ray scattering and spectroscopy, we show that this metastability is driven by a transfer of holes from chain-like charge reservoirs into the ladders. This ultrafast charge redistribution arises from the optical dressing and activation of a hopping pathway that is forbidden by symmetry at equilibrium. Relaxation back to the ground state is hence suppressed after the pump coherence dissipates. Our findings highlight how dressing materials with electromagnetic fields can dynamically activate terms in the electronic Hamiltonian, and provide a rational design strategy for nonequilibrium phases of matter.}



Ultrafast laser pulses have advanced the frontier of quantum materials research, enabling the creation of dynamical states with emergent properties and functionalities \cite{Basov2017towards}. Strong optical fields can hybridize with solids or selectively excite their microscopic degrees of freedom, leading to remarkable phenomena such as photoinduced topological \cite{wang2013observation, mciver2020light}, magnetic \cite{shin2018phonon, disa2020polarizing}, and superconducting \cite{fausti2011light,mitrano2016possible} phases. However, the transient nature of these phenomena --- often limited to the duration of the optical field or decaying shortly thereafter --- prevents their use in functional applications.

Sometimes, dynamical responses result in metastable or ``hidden'' phases of matter. These long-lived states are rare and form due to dynamical bottlenecks in their relaxation back to equilibrium. Photoexcited materials can become trapped in an intermediate state due to energy barriers created by structural effects, phase separation, and domain nucleation and growth processes \cite{Koshihara1990photoinduced,Kiryukhin1997xray, fiebig1998visualization,zhang2016cooperative,stojchevska2014ultrafast,Stoica2019optical,Cremin2019photoenhanced, nova2019metastable, sie2019ultrafast, disa2023photo, budden2021evidence}. Metastable states can also emerge due to topologically protected defects \cite{Vogelgesang2018phase,Zong2019evidence}, glassy behavior \cite{Gerasimenko2019quantum}, or trapping by impurities \cite{vonderlinde1974multiphoton}. These mechanisms rely on an interplay of structural and electronic degrees of freedom and are unique to the physics of each material. Progress in the creation and control of metastable nonequilibrium phases, however, requires the formulation of more general design principles. 

A possible strategy for achieving metastability in a broad range of quantum materials is to optically engineer their underlying Hamiltonian. This approach leverages the coherent optical dressing of electronic states by an incident electromagnetic field, a technique recently used to manipulate band structures \cite{wang2013observation, mciver2020light}, break electronic symmetries \cite{zhang2024light}, and modulate nonlinear optical properties \cite{shan2021giant} in various solids. While optical dressing is by itself transient, occurring only in the presence of the driving field, it can renormalize electronic interactions and act as a gate to induce lasting changes in the electronic distribution of a material \cite{Sato2019microscopic}. This raises the possibility of temporarily switching specific terms in the Hamiltonian and driving the system into a metastable excited state. 

While this approach may be broadly applicable, one-dimensional strongly correlated materials are particularly suited to test it due to their inherent tendency toward metastability. First, electronic relaxation in these systems is constrained by dimensionality, as charge motion is restricted to one direction. Second, partial or complete spin-charge separation suppresses the decay of noneequilibrium electronic distributions via spin fluctuations \cite{Lenarcic2015exciton}. Third, many-body interactions further slow relaxation by increasing the energy cost of scattering processes \cite{Mitrano2014pressure} and creating symmetry-protected dark states \cite{ono2004linear}. These features open the door to the realization of previously unobserved metastable or hidden states driven by purely electronic mechanisms.

Here, we observe electronic metastability in the quasi-one-dimensional cuprate ladder Sr$_{14}$Cu$_{24}$O$_{41}$. We optically induce a nonequilibrium metastable state that persists for at least tens of nanoseconds. Time-resolved terahertz reflectivity measurements, combined with ultrafast resonant x-ray scattering and spectroscopy, reveal that this metastable state involves hole transfer from chain-like charge reservoirs into the ladders. At equilibrium, symmetry constraints suppress hopping between the two structural subunits, effectively decoupling them. The pump laser optically dresses the Zhang-Rice singlet states and transiently breaks their symmetry, thus enabling hole tunneling between chains and ladders. Once the external field is removed, the symmetry is restored, trapping the holes in their new configuration. Our findings demonstrate electronic metastability through transient Hamiltonian engineering and define a general approach to realizing long-lived nonequilibrium states.

\section*{Optically-induced metastability}\vspace{-3mm}

\begin{figure}[h]
    \centering
    \includegraphics[width=1\textwidth]{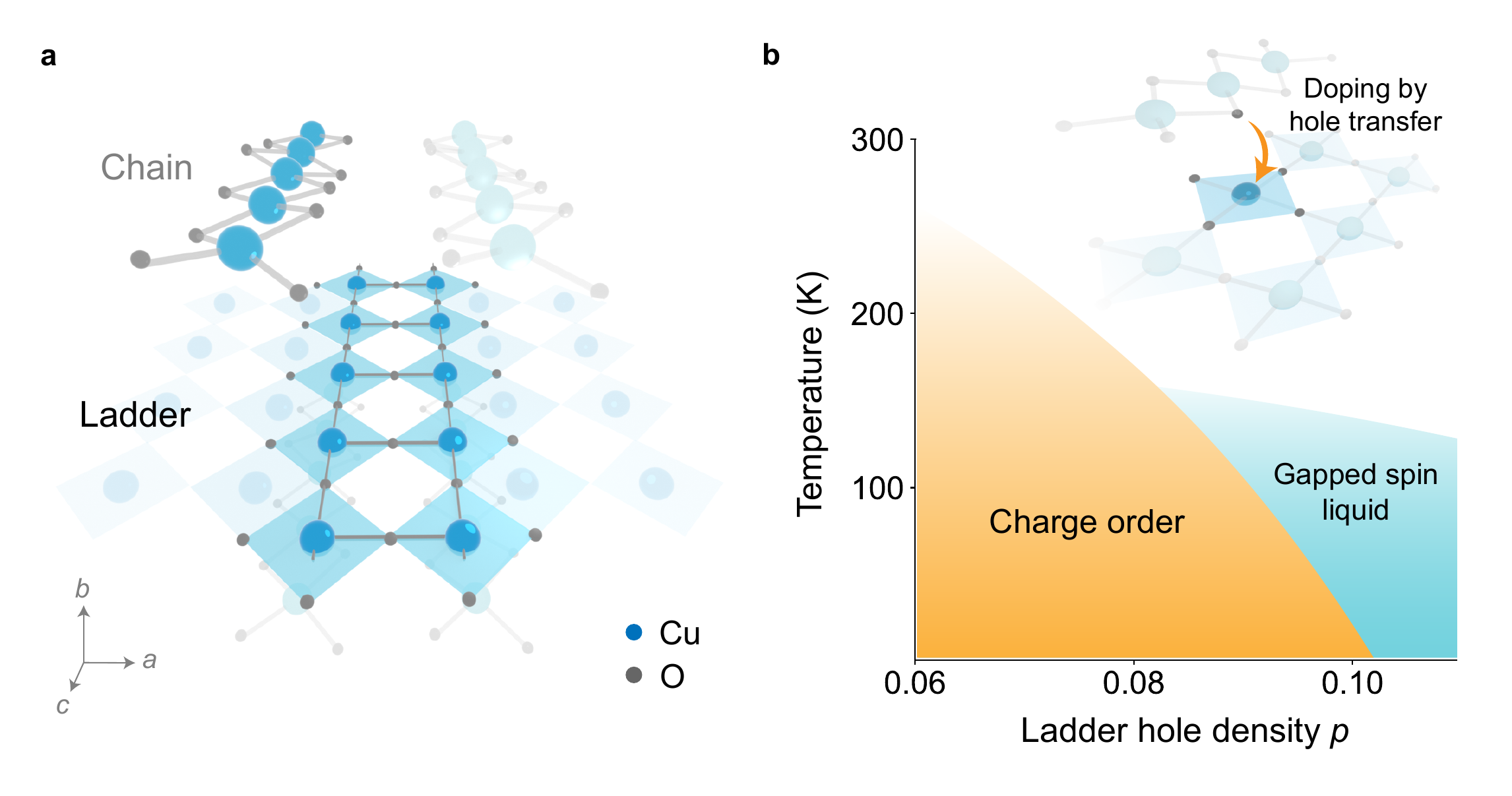}
    \caption{ {\bf Fig. ~1. Structure and electronic phases of Sr$_{14}$Cu$_{24}$O$_{41}$.} \textbf{a}, This quasi-one-dimensional compound consists of alternating ladder and chain layers with an incommensurate periodicity along the $c$ axis. The chains comprise edge-sharing CuO$_4$ squares, while the ladders feature corner-sharing CuO$_4$ plaquettes. Neighboring ladders are shifted by half a period along the $c$ axis. Intercalating Sr atoms are omitted for clarity. \textbf{b}, Schematic phase diagram \cite{Vuletic2006spinladder} of Sr$_{14}$Cu$_{24}$O$_{41}$. This compound is naturally self-doped, with most holes localized in the chains and a residual hole density of $p=0.06$ in the ladders. Isovalent Ca substitution at the Sr sites transfers holes from the chains to the ladders, increasing the ladder hole density. Upon doping, the ladders transition from a charge-ordered phase to a gapped spin liquid phase.}
    \label{fig:TRTS_intro}
\end{figure}

Sr$_{14}$Cu$_{24}$O$_{41}$ is an intrinsically self-doped charge transfer insulator with one hole per four Cu ions. Its unit cell comprises alternating layers of incommensurate chain-like and ladder-like subunits (Fig. \ref{fig:TRTS_intro}a). The chains act as hole-rich charge reservoirs \cite{osafune1997optical, nucker2000hole}, while the ladders host the remaining holes that propagate in a spin-singlet background. Below $T_\mathrm{CO}=250$ K, these carriers self-organize into a long-range charge-ordered phase \cite{abbamonte2004crystallization}. Notably, isovalent Ca substitution induces hole transfer from the chains to the ladders (Fig. \ref{fig:TRTS_intro}b), increasing the ladder hole density. This doping results in a suppression of charge order and a transition into a gapped spin liquid phase \cite{Vuletic2006spinladder}. At higher hole densities, a superconducting phase is stabilized under moderate external pressures \cite{uehara1996superconductivity, Vuletic2006spinladder}. 

\begin{figure}[h]
    \centering
    \includegraphics[width=1\textwidth]{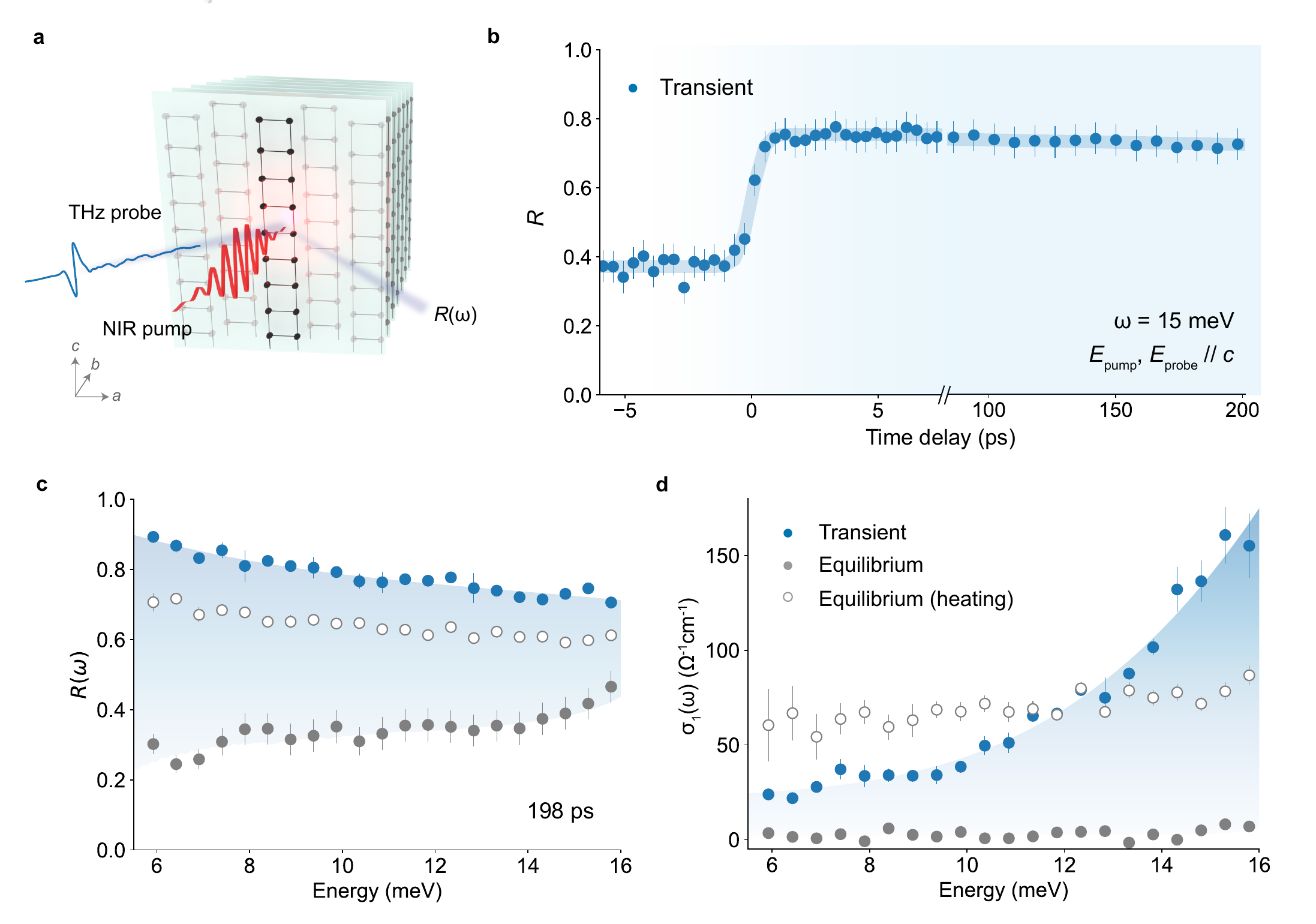}
    \caption{ {\bf Fig. ~2. Observation of light-induced metastability.} \textbf{a}, Sketch of a pump-probe experiment on the cuprate ladder Sr$_{14}$Cu$_{24}$O$_{41}$, with near-infrared (NIR) pump (1.55 eV photon energy, ~7.7 MV/cm electric field) and THz probe pulses in reflection geometry. \textbf{b}, Time-dependent reflectivity $R(\omega)$ along the $c$ direction at a representative energy (blue symbols), featuring a sharp increase and a transition into a metastable electronic state. The blue line is a fit to the data. \textbf{c}, Reflectivity $R(\omega)$ and \textbf{d}, optical conductivity $\sigma_1(\omega)$ of the metastable state ($t=198$ ps). Nonequilibrium optical properties (blue circles) are measured at 100 K. Equilibrium data are measured at 100 K (grey circles) and 250 K (open grey circles, labeled `heating'). Blue shaded areas highlight photoinduced spectral changes. Error bars are the standard deviation of 114 independent scans.}
    \label{fig:TRTS_optics}
\end{figure}

In our experiments, we use intense near-infrared pulses to drive Sr$_{14}$Cu$_{24}$O$_{41}$ single crystals (see Methods for sample characterization). Pump pulses, tuned just below the charge transfer gap energy (1.55 eV, 35 fs pulse duration), are polarized along the $c$ axis to excite Cu-O transitions along the ladder legs with peak fields up to 7.7 MV/cm. We track the pump-induced changes in low-energy optical properties using delayed quasi-single-cycle THz pulses reflected from the photoexcited samples (Fig. \ref{fig:TRTS_optics}a, also see Methods and SI Section 1). Upon exciting the sample in the charge ordered phase (100 K), we observe a reflectivity enhancement with a sub-picosecond rise time, consistent with previous reports \cite{fukaya2015ultrafast}, which unexpectedly persists for several nanoseconds (Fig. \ref{fig:TRTS_optics}b). This long-lived state is nonthermal, as indicated by the transient reflectivity and optical conductivity $\sigma_1(\omega)$ in Fig. \ref{fig:TRTS_optics}c-d, which suggests a reduction of the charge gap rather than the filling observed across the temperature-dependent charge order transition at equilibrium (Fig. S1e). These spectral changes resemble the effects of doping holes into the ladder through Ca substitution, which gradually suppresses the charge-ordered phase by closing its gap \cite{vuletic2003suppression} (see S1 Section 1 and Fig. S2). This points to a metastable enhancement of hole density in the ladders.

\section*{X-ray evidence of metastable hole transfer}\vspace{-3mm}

\begin{figure}[h]
    \centering
    \includegraphics[width=\textwidth]{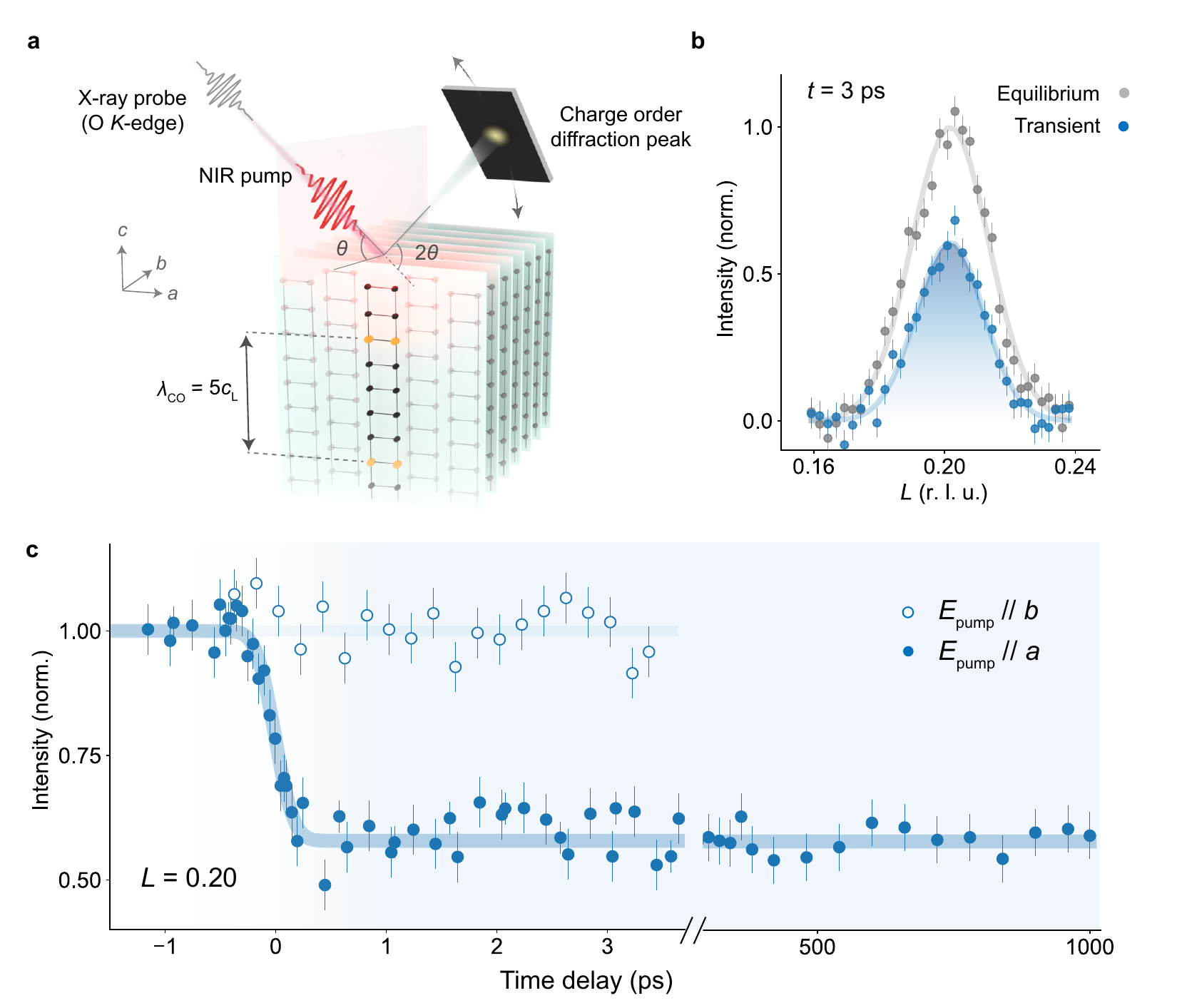} 
    \caption{ {\bf Fig. ~3. Metastable charge order suppression. } \textbf{a}, Sketch of the time-resolved x-ray diffraction experiment, with x-ray probe pulses resonant with the O $K$-edge ($\hbar\omega$ = 529 eV). The charge ordered phase with $\lambda_{\mathrm{CO}}=5c_L$ is shown schematically. \textbf{b}, Equilibrium (light grey) and transient (blue) charge order diffraction peak of the ladder. \textbf{c}, The light-induced charge order suppression is metastable up to nanosecond timescales. Error bars are the standard deviation of the signal at negative time delays. The shaded regions indicate the onset of electronic metastability. There is no peak suppression when the pump is polarized normal to the ladder plane ($E_{\mathrm{pump}}\parallel b$).}
    \label{fig:charge_order}
\end{figure}

To determine the microscopic character of the observed metastability, we interrogate the photoexcited state with a combination of ultrafast resonant x-ray techniques, each addressing a distinct observable. We first examine the nonequilibrium charge order modulation with resonant x-ray diffraction at the O $K$-edge (Fig. \ref{fig:charge_order}a). Below $T_\mathrm{CO}$ = 250~K, ladder holes form a commensurate density wave with periodicity 5$c_{\mathrm{L}}$ ($c_{\mathrm{L}}$: lattice parameter of ladder plaquettes) without detectable structural distortions \cite{abbamonte2004crystallization}. At resonance with the mobile holes, we observe an intense Bragg peak corresponding to the charge modulation at $q_{\mathrm{CO}}$ = (0, 0, 0.2) r.l.u. (Fig. \ref{fig:charge_order}b, see SI Section 2). Upon pumping along the ladder rungs, we observe a sudden reduction in diffraction intensity, consistent with a partial suppression of the charge ordered phase. This suppression persists unchanged up to 1 ns (Fig. \ref{fig:charge_order}c), indicating a long-lived photoexcited state with a lifetime exceeding tens of nanoseconds at all measured fluences (Fig. S4). This behavior contrasts sharply with that of two-dimensional cuprates, where partially suppressed charge order recovers within picoseconds \cite{mitrano2019ultrafast}. Moreover, the absence of peak broadening is incompatible with pump-induced disordering, contrary to melting involving topological defects \cite{mitrano2019ultrafast,Zong2019evidence}. Notably, the charge order suppression occurs exclusively upon pumping within the ladder plane (Fig. \ref{fig:charge_order}c), ruling out the involvement of ladder-chain dipolar excitations. The charge order suppression closely correlates with the long-lived changes in the optical properties, confirming that the charge order correlations are weaker in the nonequilibrium state.

\begin{figure}[h] 
    \centering
    \includegraphics[width=1\textwidth]{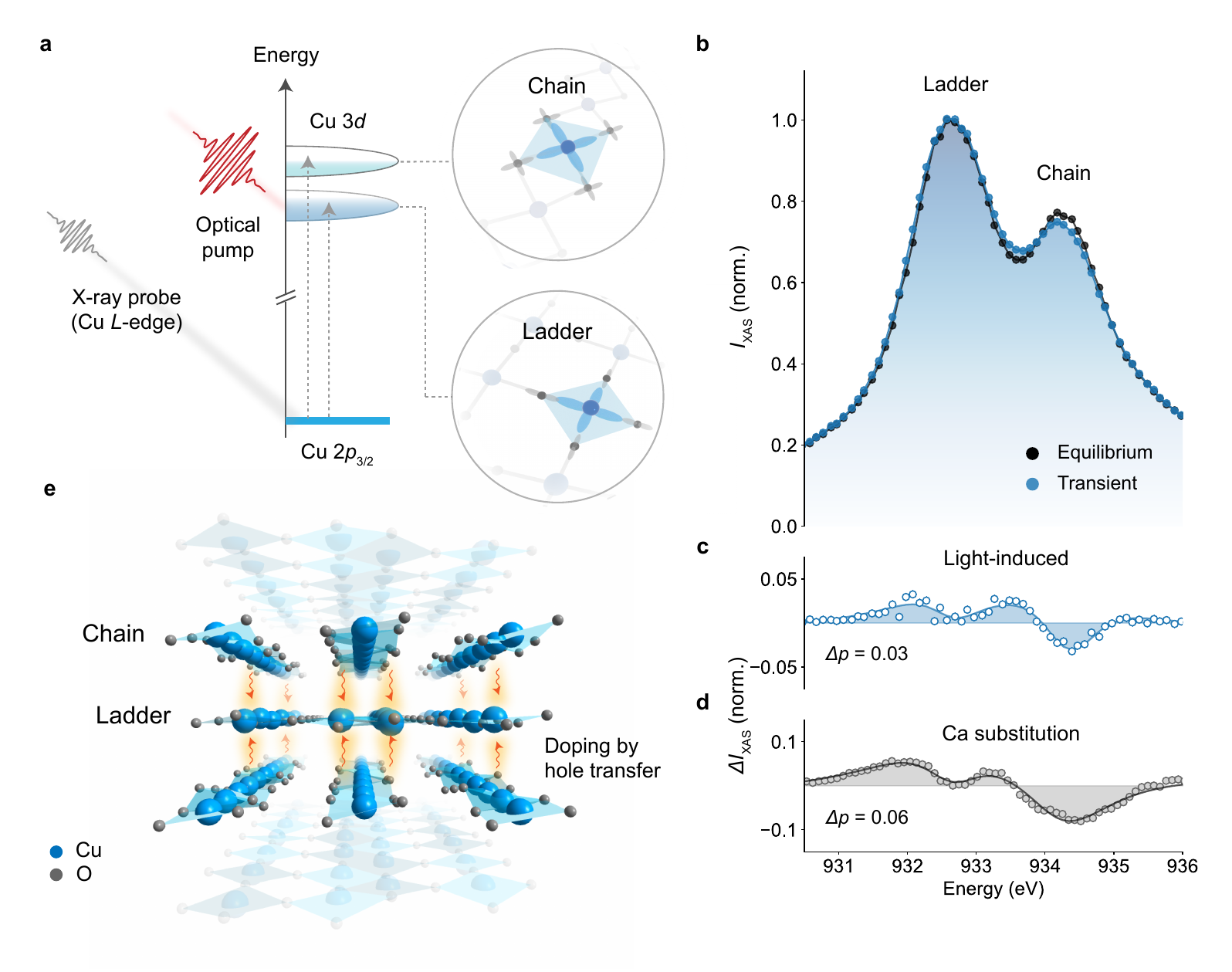}
    \caption{ {\bf Fig. ~4. Spectroscopic evidence of metastable chain-to-ladder hole transfer.} \textbf{a}, Schematic of the Cu $L$-edge time-resolved x-ray absorption spectroscopy experiment. The different chemical environments of the Cu atoms on the corner-sharing ladder and edge-sharing chain result in a double-peaked spectrum. \textbf{b}, Equilibrium (black) and transient (blue) Cu $L$$_3$-edge x-ray absorption spectra (XAS) at pump-probe delay $t$ = 3 ps. The two peaks correspond to the ladder (left) and chain (right). \textbf{c}, Differential XAS intensity of light-driven Sr$_{14}$Cu$_{24}$O$_{41}$ [$I_{\mathrm{XAS}}$($t=3$ ps) - $I_{\mathrm{XAS}}$($t<0$)]. The pump is polarized along the legs with $E_{\mathrm{pump}}\sim7.7$ MV/cm. \textbf{d}, Equilibrium XAS intensity change due to the chain-to-ladder hole transfer induced by Ca substitution in Sr$_{14-x}$Ca$_x$Cu$_{24}$O$_{41}$ (extracted from Ref. \cite{nucker2000hole}). We take the difference between the $x$ = 0 and $x$ = 11.5 compositions, corresponding to a charge transfer of $\Delta p$ = 0.06 holes/Cu$_L$. (Cu$_L$: on-ladder Cu site). \textbf{e}, Sketch of the pump-induced chain-to-ladder hole transfer.}
    \label{fig:trXAS}
\end{figure}

Next, we present a direct measurement of the valence hole distribution with time-resolved x-ray absorption spectroscopy (trXAS), which constitutes the key observation of this work. We tune the x-rays at resonance with the Cu $L_3$-edge, where the absorption spectrum features two peaks reflecting the different local bonding of Cu atoms \cite{huang2013determination} (Fig. \ref{fig:trXAS}a). The peaks at 932.5 eV and 934.4 eV each feature contributions primarily from the corner-sharing ladders and edge-sharing chains, respectively (see Fig. \ref{fig:trXAS}b, and SI Section 3, Figs. S5-7). At equilibrium, the Cu $L$-edge XAS spectrum is sensitive to changes in the hole distribution and undergoes a reshaping due to chain-to-ladder hole transfer induced by isoelectronic Ca substitution \cite{nucker2000hole, huang2013determination} (see SI Section 3, Fig. S5). The trXAS spectrum of the metastable state simultaneously shows a suppression of the chain peak and an enhancement at the shoulders of the ladder resonance (see Fig. \ref{fig:trXAS}c). The differential nonequilibrium response (Fig. \ref{fig:trXAS}c) may be compared with the effect of Ca substitution at equilibrium (Fig. \ref{fig:trXAS}d) corresponding to a chain-to-ladder hole transfer of $\Delta p$ = 0.06 holes/Cu$_{\mathrm{L}}$ \cite{nucker2000hole} (Cu$_{\mathrm{L}}$: Cu atoms on the ladder). The differential XAS changes are remarkably similar and indicate a pump-induced hole transfer from the chain reservoirs to the ladders (Fig. \ref{fig:trXAS}e). Since the amplitude of the spectral reshaping scales linearly with the transferred hole density \cite{huang2013determination} (see SI Section 3, Fig. S6), we can quantify a nonequilibrium chain-to-ladder hole transfer of $\Delta p$ = 0.03 holes/Cu$_{\mathrm{L}}$. The chain-to-ladder hole transfer is further corroborated by our trXAS measurements at the O $K$-edge (see SI Section 4, Figs. S9-10). Finally, the time-delay dependence of the trXAS spectra (see Fig. S8a) confirms that after an initial fast relaxation, the hole transfer is long-lived, mirroring the optical gap closure and charge order suppression. 

\begin{figure}[h] 
    \centering
    \includegraphics[width=\textwidth]{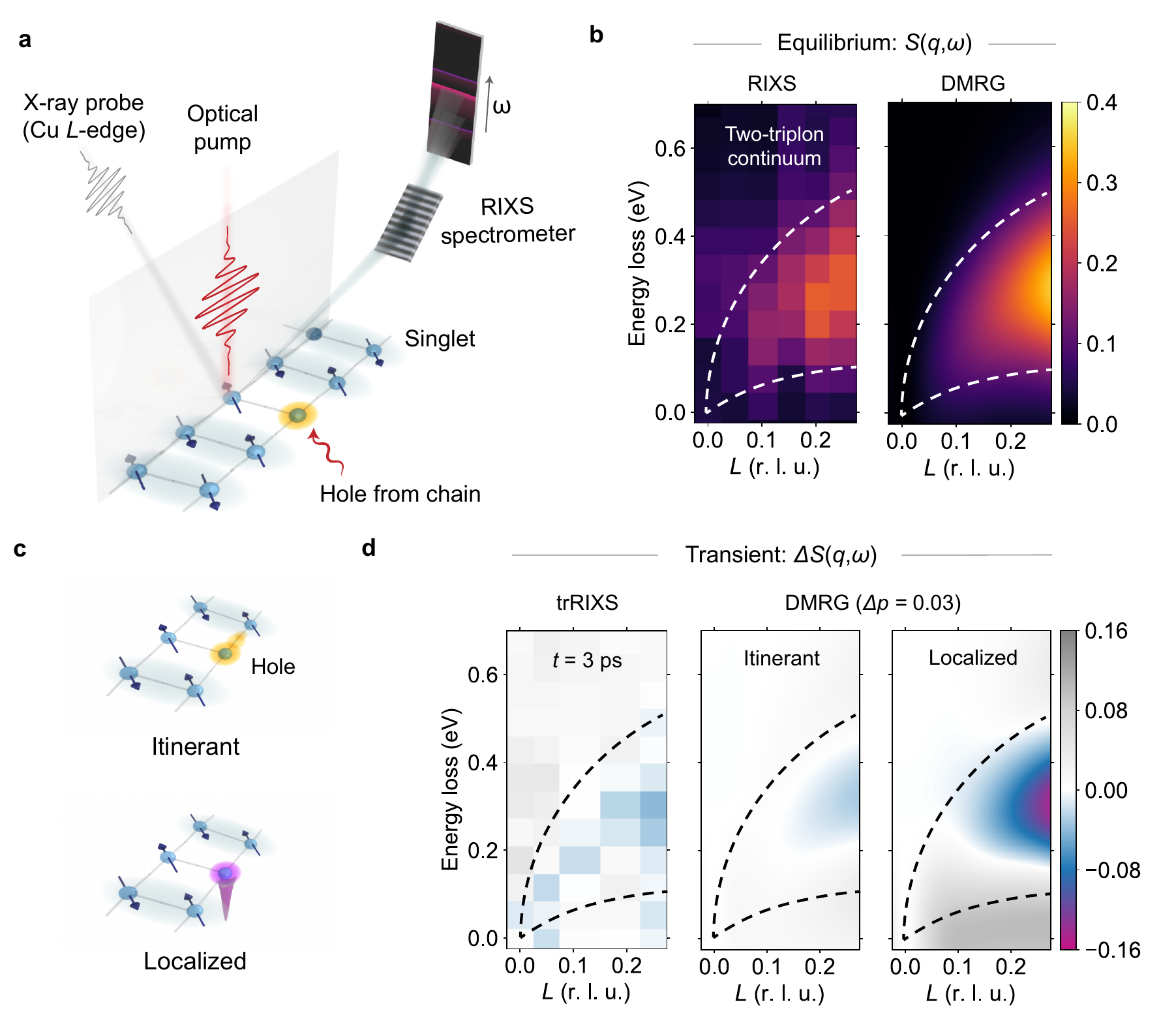} 
    \caption{ {\bf Fig. ~5. Transient magnetic excitation spectrum and onset of electronic metastability.} \textbf{a}, Sketch of the Cu $L_3$-edge time-resolved resonant inelastic x-ray scattering (trRIXS) experiment. The grey clouds denote spin singlets, and the yellow sphere denotes a hole transferred from the chain to the ladder due to the pump excitation. \textbf{b}, Intensity map of the equilibrium dynamical spin structure factor $S(q,\omega)$ measured by RIXS and calculated using DMRG. The dominant feature is a two-triplon continuum. Its boundaries are indicated by dashed lines as a guide to the eye. \textbf{c}, Sketches of itinerant (top) and localized (bottom) holes. \textbf{d}, Differential intensity map of the dynamical spin structure factor  $S(q,\omega)$ at pump-probe delay $t$ = 3 ps, as measured by trRIXS (left) and calculated using DMRG for itinerant (center) and localized (right) quasi-static hole doping $\Delta p$ = 0.03. The trRIXS data are consistent with itinerant hole doping. The RIXS energy resolution is 260 meV.}
    \label{fig:trRIXS}
\end{figure}

Metastability could potentially arise from hole localization within the ladders, which would prevent the carriers from returning to the chains. To determine whether the transferred holes in the ladders exhibit localized or itinerant character, we leverage the sensitivity of magnetic excitations to the doped carriers. We measure the magnetic excitation spectrum of the ladders with time-resolved resonant inelastic x-ray scattering (trRIXS) at the Cu $L_3$-edge (Fig. \ref{fig:trRIXS}a). Undoped ladders with isotropic couplings naturally form spin singlets \cite{dagotto1992superconductivity}. These give rise to dispersive singlet-to-triplet transitions (`triplons') as elementary magnetic excitations (see SI Section 5, Fig. S11), which appear as a two-triplon continuum at our measured wave vectors. In Fig. \ref{fig:trRIXS}b, we present the equilibrium dynamical spin structure factor $S(q,\omega)$ extracted from our RIXS data and calculated via density matrix renormalization group (DMRG) (see Methods and SI Section 5, Figs. S12-15 for more details). The experimental intensity map reveals a well-defined two-triplon continuum with positive dispersion, consistent with previous studies \cite{Schlappa2009collective}. 

The introduction of holes results in distinct changes to the two-triplon continuum, dependent on whether the holes are itinerant or localized (Fig. \ref{fig:trRIXS}c). Itinerant holes are expected to marginally suppress and broaden the triplon continuum as they disrupt the singlet background \cite{Kumar2019ladders}. In contrast, localized holes additionally localize the neighboring triplons, which would cause striking, qualitative changes to the dispersion \cite{Tseng2022crossover}. Upon photoexcitation into the metastable state, we observe a suppression of the two-triplon continuum, particularly near $L$ = 0.25 r.l.u. (Fig. \ref{fig:trRIXS}d). Within the limits of our experimental resolution, we detect no changes to the shape of the two-triplon dispersion. The differential trRIXS intensity is quantitatively reproduced in our DMRG calculations by incorporating an additional 0.03 holes/Cu$_{\mathrm{L}}$ with itinerant character. In contrast, the calculated differential intensity for localized holes (see SI Section S6) shows a suppression near $L$ = 0.25 r.l.u. that is three times greater than that observed in the experimental spectra, accompanied by enhanced intensity at lower energy loss. These results indicate the itinerant nature of the ladder holes and argue against hole localization as the cause of the observed metastability

\section*{Optical activation of a symmetry-forbidden hopping}\vspace{-3mm}

\begin{figure}[h] 
    \centering
    \includegraphics[width=1\textwidth]{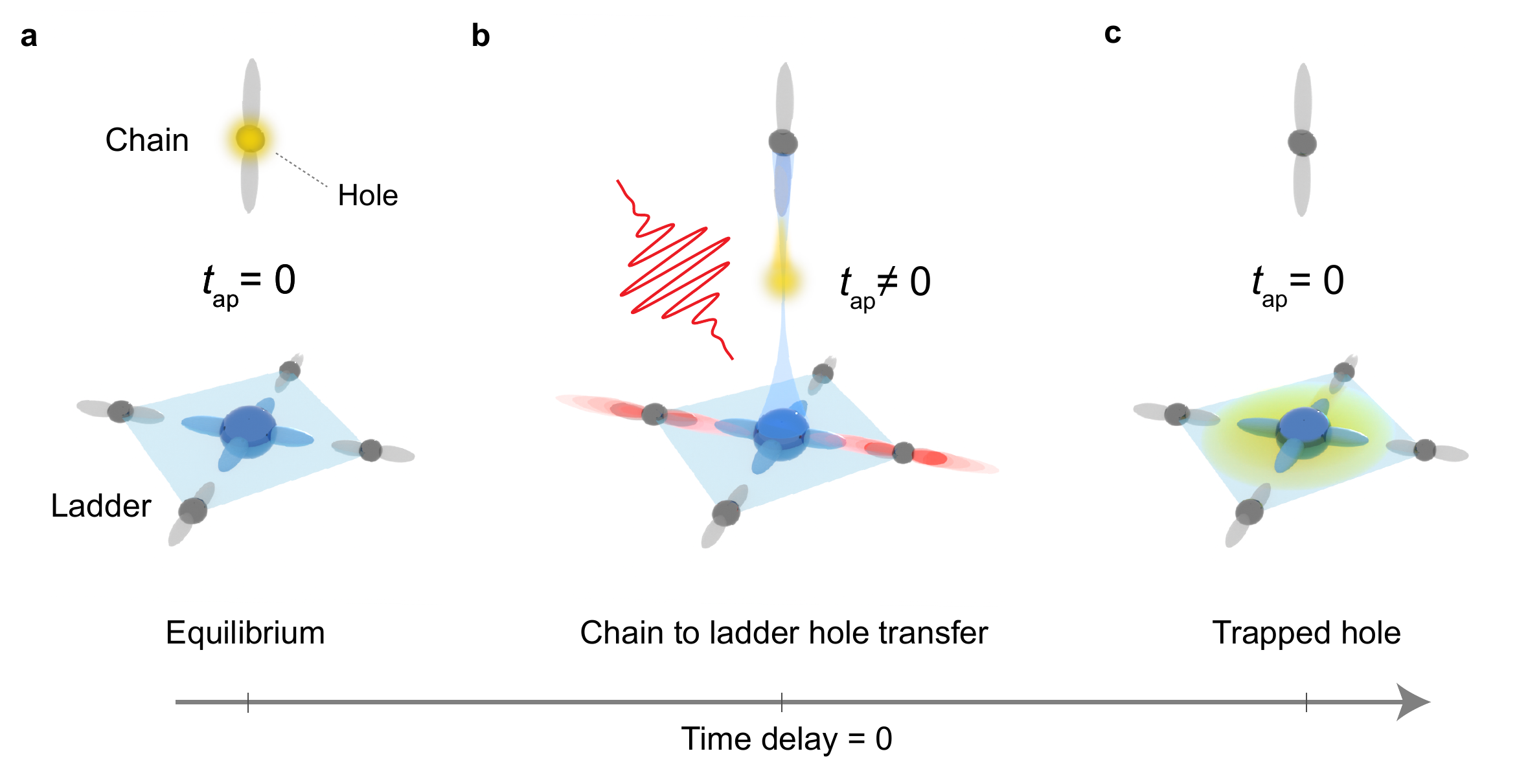} 
    \caption{ {\bf Fig. ~6. Light-induced activation of a symmetry-forbidden tunneling pathway.} \textbf{a}, Each plaquette on the ladder features O 2p (grey) and Cu 3d (blue) orbitals constituting a Zhang-Rice singlet. At equilibrium, the approximate $D_{\mathrm{4h}}$ symmetry of the Zhang-Rice singlet results in vanishing hopping $t_{\mathrm{ap}} = 0$ between the CuO$_4$ plaquette and the apical oxygen. \textbf{b}, However, this symmetry is broken when the Hamiltonian is dressed by an intense in-plane pump field $E_{\mathrm{pump}}$ (time delay = 0), resulting in a nonzero $t_{\mathrm{ap}}$. The transient activation of the symmetry-forbidden hopping $t_{\mathrm{ap}}$ leads to a chain-to-ladder hole transfer. \textbf{c}, Once the field is removed, $t_{\mathrm{ap}}$ vanishes, trapping the transferred holes in the ladder.}
    \label{fig:transient_hopping}
\end{figure}

Finally, we turn to the microscopic origin of the metastable state. The abrupt chain-to-ladder hole transfer and subsequent carrier trapping point to a transient photoinduced coupling between the two structural subunits. The coupling primarily occurs along a weak bond between copper atoms on the ladder and approximately `apical' oxygen atoms on the chain every 3-5 ladder plaquettes \cite{gotoh2003structural, deng2011structural}. At equilibrium, the direct hopping $t_\mathrm{ap}$ along this pathway is vanishingly small, resulting in an effective decoupling of holes in the chains and ladders (Fig. \ref{fig:transient_hopping}a). This is a consequence of the ladder holes forming Zhang-Rice singlet states composed of Cu $3d_{x^2-y^2}$ and surrounding O $2p_{x/y}$ orbitals on each plaquette, where the approximate $D_\mathrm{4h}$ (point group) symmetry causes hopping contributions from adjacent orbitals to cancel out (see SI Section 7). However, intense in-plane pump electric fields break this symmetry. Optical dressing of the underlying Hamiltonian unbalances hopping contributions from orbitals aligned parallel and perpendicular to the electric field, resulting in a transient, finite $t_\mathrm{ap}$ (Fig. \ref{fig:transient_hopping}b, also see SI Section S7). Such a light-activated chain-to-ladder hopping naturally explains the observed metastability. While the chains and ladders are effectively decoupled at equilibrium, the light-induced enhancement of $t_{\mathrm{ap}}$ allows charge transfer between the two subsystems over the duration of the pump pulse. After the pulse, the chains and ladders are once again decoupled, and the light-induced itinerant holes in the ladder become symmetry-protected against relaxation (Fig. \ref{fig:transient_hopping}c). 

This coherent dressing mechanism predicts a light-induced hole transfer from the chains to the ladders, with $\Delta p \propto E_\mathrm{pump}^2$ (see Fig. S19b). This is consistent with the sign and field dependence of the differential ladder and chain XAS intensities in the metastable state (Fig. S8b). The experimentally observed scaling with $E_{\mathrm{pump}}^2$ also rules out higher-order dipole-type transitions across the Mott gap, which would instead scale with $E_{\mathrm{pump}}^4$. Finally, contrary to a dipole-type chain-ladder optical transition, such a symmetry-protected charge-transfer mechanism would have a vanishing response for an out-of-plane-polarized pump ($E_{\mathrm{pump}} \parallel b$) (see SI Section 7), in agreement with the experimental polarization dependence of the light-induced charge order suppression (Fig. \ref{fig:charge_order}c). 

Given that metastability often relies on cooperative structural responses, we finally consider the role of photoinduced lattice changes. One could posit a structural distortion that brings chains and ladders closer and facilitates the transfer of holes between subunits. At equilibrium, such a transfer via Ca substitution indeed coincides with a reduction of both $b$- and $c$-axis lattice constants \cite{deng2011structural}. However, our O $K$-edge trXAS measurements reveal spectral changes consistent with chain-to-ladder hole transfer, but without the blue shift associated with $b$ axis structural distortions of ladder oxygens (see SI Section S4). Moreover, a contraction of the $c$-axis, similar to that induced by Ca substitution, would increase the charge order scattering momentum by $\sim 4-6\cdot 10^{-3}$ r.l.u., as the wavevector is doping-independent until abruptly transitioning to very different commensurability at high calcium concentrations \cite{Rusydi2006quantum}. Within our experimental resolution, we do not observe such an increase. Finally, the time evolution of our observables, namely a sudden change upon photoexcitation and the absence of measurable relaxation, contrasts sharply with the typical structural response of other photoexcited cuprates. Their lattice response typically features a picosecond-long saturation build-up \cite{Gedik2007nonequilibrium,Mansart2013temperature} and a relaxation time well below nanosecond timescales \cite{Gedik2007nonequilibrium,Mankowsky2014nonlinear}. In our experiments, holes are transferred on a resolution-limited timescale (Fig. S8), indicating they are unlikely driven by a structural response of the type observed in other copper oxides. While more subtle structural changes may be resolved with further experimentation, our evidence strongly supports a photoinduced coupling of chains and ladders via apical oxygen hopping as the primary and simplest driver of the observed metastability

\section*{Outlook}\vspace{-3mm}

Our findings establish Hamiltonian engineering through optical dressing as a powerful strategy to create metastable nonequilibrium states in quantum materials. While one-dimensional systems are fertile ground for metastability, this approach can be extended to other layered materials such as 2D oxide superconductors \cite{keimer2015quantum}, hybrid perovskites \cite{shi2023rapid}, and van der Waals heterostructures \cite{jin2018ultrafast}. In these systems, selective activation of symmetry-forbidden hopping terms could allow for precise doping of specific bands beyond the limits of chemical substitution, and enable light-driven steering across correlated electronic phases. Additionally, this coherent control protocol could be used to dynamically modulate the interlayer charge distribution in tailored heterostructures for optoelectronic applications at slower timescales. Finally, upon cooling, long-lived nonequilibrium charge distributions could give rise to new light-induced phenomena, including spin and orbital ordering, excitonic condensation, and $\eta$-pairing superconductivity. 
\clearpage

\clearpage

\clearpage

\clearpage

\clearpage

\clearpage

\bibliography{sn-bibliography}
\clearpage

\section*{Methods}

\subsection*{Crystal synthesis and characterization}\vspace{-4mm}
High-quality single crystals of Sr$_{14}$Cu$_{24}$O$_{41}$ were grown using a modified traveling solvent floating zone (TSFZ) technique. We started by synthesizing polycrystalline Sr$_{14}$Cu$_{24}$O$_{41}$ via a solid-state reaction, and subsequently used it as a feed material rod. Choosing CuO as the flux, we prepared the flux rod (seed) for the TSFZ growth by mixing Sr$_{14}$Cu$_{24}$O$_{41}$ and CuO powders with a mass ratio of Sr$_{14}$Cu$_{24}$O$_{41}$:CuO = 1:0.0163. To ensure stable growth, we tuned the feed speed of the feed rod within 0.81-2.2 mm/hour, and set the growth speed to 0.81 mm/hour. We rotated both rods in mutually opposite directions at 30 rpm and finally obtained a crystalline rod (2 cm length, 4 mm diameter). We confirmed crystal quality and lattice structure via x-ray diffraction measurements, finding good agreement with previous reports \cite{vanishri2009crystal}. The crystal lattice parameters are $a$ = 11.47 \AA, $b$ = 13.35 \AA, and $c$ = 7$c_\mathrm{L}$ = 10$c_\mathrm{C}$ = 27.46 \AA, where `L' and `C' subscripts indicate ladder and chain subunits, respectively. 

\subsection*{trTDTS measurements}\vspace{-4mm}
The optical setup for our time-resolved time-domain THz spectroscopy experiments (trTDTS) was seeded by a Ti:sapphire regenerative amplifier (800 nm wavelength, 35 fs  pulse duration , 2 kHz repetition rate). We split the beam into three branches. The first branch (0.25 mJ) was used for the pump excitation. The second branch (0.5 mJ pulse energy) was used to generate quasi-single-cycle THz probe pulses with spectral components between 1 and 5 THz (4-20 meV) via optical rectification in a 0.2-mm-thick \textlangle 110\textrangle~GaP crystal. The third branch (0.05 mJ) was used as an optical gate for the electro-optic sampling (EOS) of the probe pulses reflected from the sample surface. The pump-delay was controlled by a mechanical delay stage on the pump beam.

\textbf{Single-shot THz detection}: We used echelon-based single-shot detection \cite{gao2022high} to measure equilibrium and transient THz fields. The optical gate pulse was routed through an echelon mirror, resulting in a temporally offset array (spacing 20 fs) of pulselets spanning a 10-ps time window. We focused the pulselets onto a 0.2-mm-thick \textlangle 110\textrangle~GaP crystal to perform single-shot EOS of the probe pulses reflected from the sample surface. The transmitted beam was split into perpendicularly-polarized components using a quarter-waveplate and a Wollaston prism, and each component was imaged on a linear array detector (Synertronics Glaz LineScan with a Hamamatsu S11637-2048Q sensor). The final THz waveform was obtained by calibrating and subtracting the two images, resulting in $E(t_\mathrm{EOS})$, where $t_\mathrm{EOS}$ is the THz EOS time delay. For our pump-probe experiments, we measured this as a function of the pump delay $t$ to obtain a 2D map $E(t_\mathrm{EOS}, t)$. We interpolated and appropriately translated the measured $E(t_\mathrm{EOS}, t)$ along the $t$ axis to ensure that the transient waveform at each t was measured with the gate and pump pulses at the same $t_\mathrm{EOS}$ \cite{hu2014optically}. 

\textbf{trTDTS measurements}: We conducted our measurements in a reflection geometry on freshly-cleaved $ac$ surfaces of Sr$_{14}$Cu$_{24}$O$_{41}$ single crystals. The THz probe beam was S-polarized ($E \parallel c$) and focused onto the sample at an angle of incidence of 60$^\mathrm{o}$, using an off-axis parabolic mirror. The reflected light was focused onto the GaP crystal for the EOS. The pump beam was focused onto the sample at normal incidence with a spot size of 1400 $\mu$m, while the probe spot size was 600 $\mu$m. We mechanically chopped the pump at twice the repetition rate of the probe beam. The linear array detectors were connected to a digitizer synchronized to the mechanical choppers which sorted the detected THz waveforms into `pump on' and `pump off' conditions. This allowed us to simultaneously acquire equilibrium and transient THz waveforms, and eliminate any possible artifacts due to long-term drift and residual pump scattering.

\textbf{Reconstruction of transient optical conductivity}: The equilibrium electric field $E_0(t_\mathrm{EOS})$ and differential transient electric field $\Delta E(t_\mathrm{EOS}, t)$ were independently Fourier transformed to obtain the complex Fresnel reflection coefficient $\tilde{r}(\omega, t)$ using the expression

\begin{equation}
\Delta \tilde{E}/\tilde{E}_0 = (\tilde{r}(\omega, t) - \tilde{r}_0(\omega))/\tilde{r}_0(\omega),
\end{equation}

where $\tilde{r}_0$ is the reflection coefficient calculated from the equilibrium optical response (see SI Section 1, Fig. S1). From this, we evaluated the transient refractive index $\tilde{n}(\omega, t)$. Since the probe penetration depth (45-50 $\mu$m) is a factor of 150 larger than that of the pump (0.32 $\mu$m), we employed the thin film approximation, wherein it is assumed that a thin layer at the sample surface is homogeneously photoexcited, while the bulk remains unperturbed \cite{fausti2011light,hunt2015manipulating}. The analytical expression for the transient change to the optical conductivity in the photoexcited volume is given by:

\begin{equation}
\Delta\tilde{\sigma}(\omega, t) = \left(\frac{1}{377 \times \delta}\right) \frac{ \frac{\Delta\tilde{E}(\omega, t)}{\tilde{E}_0(\omega)} \left(\tilde{n}^2(\omega, t) - 1\right)}{\frac{\Delta\tilde{E}(\omega, t)}{\tilde{E}_0(\omega)} \left[\cos\theta_0 - \sqrt{\tilde{n}^2(\omega, t) - \sin^2\theta_0}\right] + 2\cos\theta_0},
\end{equation}

where $\delta$ is the pump penetration depth and $\theta_0$ is the probe angle of incidence.

\subsection*{trXAS and trRIXS measurements}\vspace{-4mm}
We conducted time-resolved x-ray absorption spectroscopy (trXAS) and resonant inelastic x-ray scattering (trRIXS) measurements at the Furka endstation of Athos beamline at SwissFel, Paul Scherrer Institut \cite{Abela2019swissfel}. The repetition rate was 100 Hz. The x-ray beam was horizontally polarized and focused to a spot size of 600 $\mu$m (H) $\times$ 10 $\mu$m (V). Shot-to-shot x-ray intensity fluctuations were recorded with an avalanche photodiode (APD) and used to normalize the signals. We used 800 nm (1.55 eV), 100-fs-long pump pulses, which were focused down to a diameter of 1300 $\mu$m to achieve fluences up to 8 mJ/cm$^2$ (approximately 8 MV/cm). The pump penetration depth (0.32 $\mu$m) exceeded that of the soft x-ray probe at the Cu $L$-edge exceeded that of the soft x-ray probe ($\sim$0.2-0.3 $\mu$m) at all measured angles of incidence, hence resulting in a homogeneously excited probed sample volume. We cleaved the sample along the $b$ axis in situ, and mounted it with the $bc$ axes in the scattering plane. The temperature was fixed to 100 K for all measurements. 

\textbf{trXAS measurements}: We acquired spectra in fluorescence-yield mode at the Cu $L$$_3$-edge, with the x-ray beam near normal incidence and detected by an avalanche photodiode at 2$\theta$ = 78$^\mathrm{o}$. We acquired 2000 pulses at each time delay for the time-dependent XAS intensity traces and 1000 pulses at each monochromator energy for the trXAS spectra. 

\textbf{trRIXS measurements}: We acquired RIXS spectra using incident x-rays resonant with the Cu $L_3$ peak at 932.6 eV. Our measurements were performed with the scattering angle 2$\theta$ fixed at 130$^\mathrm{o}$ and the incident angle $\theta$ varied from 65$^\mathrm{o}$ to 93$^\mathrm{o}$, corresponding to momentum transfers of $L$ = 0 to 0.25 r. l. u. (defined in units of 2$\pi/c_\mathrm{L}$). Given the low-dimensional nature of the spin fluctuations, we neglected dispersion along the $K$ direction. The RIXS spectrometer had a combined energy resolution of 260 meV. We collected each trRIXS spectrum by acquiring 30,000 shots each with the pump laser on and off, and averaging over 8 scans each, measured in an alternating manner to monitor and eliminate artifacts due to slow drifts. The raw spectra, subtraction of the elastic line, and extraction of the dynamical spin structure factor $S(q, \omega)$ are shown in SI Section S5, Figs. S10-13.

\subsection*{trXRD measurements}\vspace{-4mm}
We conducted time-resolved x-ray diffraction (trXRD) resonant with the O $K$-edge at 528.6 eV at the RSXS endstation of the Pohang Accelerator Laboratory x-ray Free Electron Laser running at a repetition rate of 30 Hz. The x-ray pulses were horizontally polarized and focused to a spot size of 160 $\mu$m (H) $\times$ 295 $\mu$m (V). We recorded shot-to-shot intensity fluctuations using a gas monitor detector and used them to normalize the signals. The pump pulses had a duration of 100 fs, and were focused to a spot size of 585 $\mu$m (H) $\times$ 549 $\mu$m (V), with fluences up to 6 mJ/cm$^2$. The pump penetration depth (0.32 $\mu$m) is comparable to that of the O $K$-edge x-ray probe (0.28 $\mu$m). Measurements were done on a polished sample, with the surface normal oriented parallel to the $c$ axis with a miscut of 10$^\mathrm{o}$. We detected the CO peak at $q$ = [0, 0, 0.2] by fixing the incident angle $\theta$ at 36$^\mathrm{o}$ and the scattering angle 2$\theta$ at 72$^\mathrm{o}$. Details about the subtraction of the fluorescent background are provided in the SI. We additionally conducted O $K$-edge time-resolved x-ray absorption spectroscopy measurements, outlined in SI Section 4. We acquired spectra in fluorescence-yield mode, with the x-ray beam near normal incidence and detected by an APD at 2$\theta$ = 110$^\mathrm{o}$. All measurements were performed at 100 K. 

\subsection*{Density matrix renormalization group calculations}\vspace{-4mm}
We employed DMRG to calculate the ground state $|G\rangle$ of an extended Hubbard model, described by the Hamiltonian
\begin{eqnarray}
    \mathcal{H} &=& - \sum_{jl\sigma}  t\big[c^{(l)\dagger}_{j\sigma} c^{(l)}_{j+1\sigma} + h.c.\big] -\sum_{j \sigma}  t_\perp \big[c^{(0)\dagger}_{j\sigma} c^{(1)}_{j\sigma} + h.c.\big] -\sum_{jl \sigma}  t^\prime \big[c^{(l)\dagger}_{j\sigma} c^{(1-l)}_{j+1\sigma} + h.c.\big]\nonumber\\
    &&+ U \sum_{jl} n^{(l)}_{j\uparrow} n^{(l)}_{j\downarrow} + V\sum_{j}\sum_{\sigma,\sigma^\prime}\big[ n^{(0)}_{j\sigma} n^{(1)}_{j\sigma^\prime}+\sum_{l}n^{(l)}_{j\sigma} n^{(l)}_{j+1\sigma^\prime}\big],
\end{eqnarray}
where $c^{(l)}_{j\sigma}$ ($c^{(l)\dagger}_{j\sigma}$) annihilates (creates) an electron at site $j$ on leg $l = 0,1$ with spin $\sigma = \uparrow, \downarrow$, and $n^{(l)}_{j\sigma} = c^{(l)^\dagger}_{j\sigma} c^{(l)}_{j\sigma}$ denotes the local electron density. We take the on-site Coulomb interaction $U=8t$ and an attractive NN interaction $V=-1.25t$. We use the time-dependent variational principle (TDVP)\,\cite {haegeman2011time} to simulate the time evolution of the wavefunction after a local spin excitation, giving as a unequal-time correlation function 
\begin{equation} \label{tDMRG_2leg} 
    S(q, \omega) = \int_0^{T_{\text{max}}} dt \sum_{j} \sum_{l=0,1} \left\langle G \left| \mathcal{U}(0,t)\, S^{(l)}_{j}\mathcal{U}(t,0) S^{(0)}_{j_0} \right| G \right\rangle e^{i q j}  e^{-i \omega t}\,,
\end{equation}
which corresponds to the RIXS measurements with matrix-element corrections applied. Here, $S^{(l)}_j= \big[c^{(l)\dagger}_{j
\uparrow}c^{(l)}_{j\uparrow}-c^{(l)\dagger}_{j\downarrow}c^{(l)}_{j\downarrow}\big]/2$ is the spin operator at site $j$ on leg $l$, and $\mathcal{U}(t_1,t_2)$ is the time-evolution operator. To minimize boundary effects and enforce translational symmetry, the middle site $j_0=L/2$ is fixed, and the sum in Eq.~\eqref{tDMRG_2leg} runs over all site indices $j$. In this work, we keep the maximum bond dimension $D = 1000$, with truncation error around $10^{-7}$. The time evolution has a step $\delta t= 0.05t^{-1} $ and is truncated at $T_{\text{max}} = 30$. All the DMRG results are obtained using a two-leg ladder with $L_x=64$ and broadened to match the experimental energy resolution.

\subsection*{Density functional theory calculations}\vspace{-4mm}
Density functional theory (DFT) simulations of the electronic structure and of the Cu $L$-edge XAS spectra have been carried out with \textsc{Quantum Espresso} (QE) version 7.3.1 \cite{giannozzi_quantum_2009, giannozzi_advanced_2017} via the \textsc{pw.x} and \textsc{xspectra.x} packages \cite{taillefumier_x-ray_2002, gougoussis_first-principles_2009, gougoussis_intrinsic_2009, bunau_projector_2013}. QE implements DFT within the pseudopotential (PP) and planewaves (PW) approach. For our simulations, we used ultrasoft PPs with gauge invariant projector augment wave (GIPAW) reconstructions \cite{pickard_all-electron_2001}, non-linear core corrections, and two projectors per angular momentum channel to ensure the convergence of the XAS spectra. We imposed a kinetic energy cutoff of 50 Ry (680 eV) on the PW basis set for the Kohn-Sham wavefunctions and 500 Ry (6800 eV) for the electronic density. Electron exchange and correlation effects have been treated using the Perdew-Burke-Ernzerhof (PBE) generalised gradient approximation (GGA) \cite{perdew_generalized_1996} with Hubbard $+U$ corrections applied to the Cu $d$ (10.0 eV) and O $p$ (4.0 eV) states \cite{ilakovac_hole_2012}. We adopted the rotationally invariant DFT+$U$ approach using orthogonalised atomic orbitals as Hubbard the projectors \cite{dudarev_electron-energy-loss_1998}.
For the DFT steps the integration of the first Brillouin zone has been conducted using a $2\times 2\times 1$ regular $\Gamma$-centered \textbf{k}-point mesh. For the calculation of the XAS spectrum the \textbf{k}-point mesh is shifted by half a grid spacing along each of the Cartesian directions.

These simulation parameters and the initial atomic coordinates and magnetic ordering of the $1\times 1\times 4$ supercell have been adapted from a previous investigation of the O $K$-edge in the same compound \cite{ilakovac_hole_2012}. The stoichiometry of the system investigated was Cu$_{96}$O$_{164}$Sr$_{56}$ and the dimensions of the supercell were $a=b=8.825\ \mathrm{\AA} ,\ c=55.058\ \mathrm{\AA}$ and an angle $\angle ab = 81.146^\circ$. The crystal structure has been relaxed (using atomic orbitals as Hubbard projectors) with the Broyden-Fletcher-Goldfarb-Shanno quasi-Newton algorithm; once a force tolerance of 1.3$\times 10^{-3}$ atomic units was reached the structure was deemed to be relaxed. For the calculation of the Cu $L$-edge XAS spectrum we created isolated ladder (Cu$_{56}$O$_{84}$) and chain (Cu$_{40}$O$_{80}$) subunits maintaining the same supercell dimensions as the parent compound. Treating the subunits in isolation enabled us to simulate the effects of the charge transfer by modulating the number of holes on each component. We applied a total charge of +4.0 $e$ and +20.0 $e$ on the ladder and chain respectively, which corresponds to the equilibrium state. For the optically pumped state, to simulate hole transfer these charges were set to +6.0 $e$ and +18.0 $e$ for the ladder and chain respectively. These charges have been chosen to match to the experimentally determined ladder hole densities of 0.06 and 0.09/Cu$_\mathrm{L}$. The XAS spectra were computed at each inequivalent Cu site within the dipole approximation, along the [0 0 1] polarisation direction using the Lanczos-Haydock recursive fraction algorithm \cite{taillefumier_x-ray_2002}. A Lorentzian broadening of 300 meV is applied to the final spectrum to account for core-hole broadening effects.

\section*{Data availability}\vspace{-4mm}
The data that support the findings of this study are present in the paper and/or in the Supplementary Information. Additional data related to the paper are available from the corresponding authors upon reasonable request.

\section*{Acknowledgements}\vspace{-4mm}
We thank C. Bernhard, A. Cavalleri, S. Chattopadhyay, R. Comin, E. Demler, M. Eckstein, T. Giamarchi, V. Ilakovac, and S. Johnston for insightful discussions. Experimental part of this work was primarily supported by the U.S. Department of Energy, Office of Basic Energy Sciences, Early Career Award Program, under Award No. DE-SC0022883. Theoretical part of the work (L.X., Z.S., and Y.W.) was supported by the Air Force Office of Scientific Research Young Investigator Program under Grant No. FA9550-23-1-0153. Work performed at Brookhaven National Laboratory was supported by the U.S. Department of Energy (DOE), Division of Materials Science, under Contract No. DE-SC0012704. B. Lee and H.J. were supported by the National Research Foundation of Korea (MSIT), Grant No. 2022M3H4A1A04074153 and 2020M3H4A2084417. M.C. acknowledges support from the European Union (ERC, DELIGHT, 101052708). We acknowledge the Paul Scherrer Institut, Villigen, Switzerland, for the provision of beamtime at the Furka beamline of the SwissFEL. The work at the PAL-XFEL was performed at the RSXS endstation (Proposal No. 2023-1st-SSS-002), funded by the Korea government (MSIT). The single crystal growth work was performed at the Pennsylvania State University Two-Dimensional Crystal Consortium–Materials Innovation Platform (2DCC-MIP), which is supported by NSF Cooperative Agreement No. DMR-2039351. JDE and MCB are supported by The Royal Society, Grant No.~IES/R3/223185. 
We acknowledge computational resources from ARCHER2 UK National Computing Service which was granted via HPC-CONEXS, the UK High-End Computing Consortium (EPSRC grant no. EP/X035514/1). The simulation used resources of the Frontera computing system at the Texas Advanced Computing Center.

\section*{Author contributions}\vspace{-4mm}
H.P. and M.M. conceived the project. M.M. supervised the project. H.P. and F.G. conducted the trTDTS measurements. C.H. performed the equilibrium optical spectroscopy measurements. H.P., S.T., M.P.M.D., M.M., E.S., H.U., B. Liu, E.P., and E.R. conducted the Cu $L$-edge trXAS and trRIXS measurements. H.P., S.T., B. Lee, H.C., S.Y.P., and H.J. conducted the O $K$-edge trXAS and trXRD measurements.  Yu W., S.H.L., and Z.M. synthesized the samples. H.P., W.H., and S.T. prepared and pre-characterized the samples. Yao W. developed the theoretical model for the light-activated hopping mechanism. Z.S. and H.W. performed the DMRG calculations and L.X. performed the ab initio simulations under the supervision of Yao W. J.E. and M.C. performed the DFT calculations. H.P. analyzed the data. H.P., Yao W., and M.M. wrote the manuscript with input from all authors.

\section*{Competing interests}\vspace{-4mm}
The authors declare no competing interests.

\end{document}


\title[Article Title]{Symmetry-protected electronic metastability in an optically driven cuprate ladder}


\author*[1]{\fnm{Hari} \sur{Padma}}\email{hpadmanabhan@g.harvard.edu}
\author[1]{\fnm{Filippo} \sur{Glerean}}
\author[1,2]{\fnm{Sophia} \sur{F. R. TenHuisen}}
\author[3]{\fnm{Zecheng} \sur{Shen}}
\author[4]{\fnm{Haoxin} \sur{Wang}}
\author[3]{\fnm{Luogen} \sur{Xu}}
\author[5]{\fnm{Joshua} \sur{D. Elliott}}
\author[6]{\fnm{Christopher} \sur{C. Homes}}
\author[7]{\fnm{Elizabeth} \sur{Skoropata}}
\author[7]{\fnm{Hiroki} \sur{Ueda}}
\author[7]{\fnm{Biaolong} \sur{Liu}}
\author[7]{\fnm{Eugenio} \sur{Paris}}
\author[7]{\fnm{Arnau} \sur{Romaguera}}
\author[8,9]{\fnm{Byungjune} \sur{Lee}}
\author[10]{\fnm{Wei} \sur{He}}
\author[11,12]{\fnm{Yu} \sur{Wang}}
\author[11,12]{\fnm{Seng Huat} \sur{Lee}}
\author[13]{\fnm{Hyeongi} \sur{Choi}}
\author[13]{\fnm{Sang-Youn} \sur{Park}}
\author[11,12]{\fnm{Zhiqiang} \sur{Mao}}
\author[14]{\fnm{Matteo} \sur{Calandra}}
\author[13]{\fnm{Hoyoung} \sur{Jang}}
\author[7]{\fnm{Elia} \sur{Razzoli}}
\author[10]{\fnm{Mark P. M.} \sur{Dean}}
\author*[3]{\fnm{Yao} \sur{Wang}}\email{yao.wang@emory.edu}
\author*[1]{\fnm{Matteo} \sur{Mitrano}}\email{mmitrano@g.harvard.edu}

\affil[1]{\orgdiv{Department of Physics}, \orgname{Harvard University}, \orgaddress{\city{Cambridge}, \state{MA}, \country{USA}}}

\affil[2]{\orgdiv{Department of Applied Physics}, \orgname{Harvard University}, \orgaddress{\city{Cambridge}, \state{MA}, \country{USA}}}

\affil[3]{\orgdiv{Department of Chemistry}, \orgname{Emory University}, \orgaddress{\city{Atlanta}, \state{GA}, \country{USA}}}

\affil[4]{\orgdiv{Department of Physics}, \orgname{The Chinese University of Hong Kong}, \orgaddress{\city{Hong Kong}}}

\affil[5]{\orgname{Diamond Light Source}, \orgaddress{\city{Didcot}, \state{Oxfordshire}, \country{UK}}}

\affil[6]{\orgdiv{National Synchrotron Light Source II}, \orgname{Brookhaven National Laboratory}, \orgaddress{\city{Upton}, \state{NY}, \country{USA}}}

\affil[7]{\orgname{PSI Center for Photon Science}, \orgname{Paul Scherrer Institute}, \orgaddress{\city{Villigen}, \country{Switzerland}}}

\affil[8]{\orgdiv{Department of Physics}, \orgname{Pohang University of Science and Technology}, \orgaddress{\city{Pohang}, \country{Korea}}}

\affil[9]{\orgdiv{Max Planck POSTECH/Korea Research Initiative}, \orgname{Center for Complex Phase Materials}, \orgaddress{\city{Pohang}, \country{Korea}}}

\affil[10]{\orgdiv{Condensed Matter Physics and Materials Science Department}, \orgname{Brookhaven National Laboratory}, \orgaddress{\city{Upton}, \state{NY}, \country{USA}}}

\affil[11]{\orgname{Department of Physics}, \orgname{Pennsylvania State University}, \orgaddress{\city{University Park}, \state{PA}, \country{USA}}}

\affil[12]{\orgname{2D Crystal Consortium}, \orgname{Materials Research Institute}, \orgname{Pennsylvania State University}, \orgaddress{\city{University Park}, \state{PA}, \country{USA}}}

\affil[13]{\orgname{Pohang Accelerator Laboratory (PAL)}, \orgname{Pohang University of Science and Technology}, \orgaddress{\city{Pohang}, \country{South Korea}}}

\affil[14]{\orgname{Department of Physics}, \orgname{University of Trento}, \orgaddress{\city{Povo}, \country{Italy}}}

\maketitle

\vspace{-10mm}
\section*{Supplementary Information}

\begin{multicols}{2}
\begin{enumerate}
\RaggedRight
\item \textbf{trTDTS measurements}
\item \textbf{trXRD measurements}
\item \textbf{trXAS measurements at Cu $L$-edge}
\item \textbf{trXAS measurements at O $K$-edge}
\item \textbf{trRIXS measurements}
\item \textbf{DMRG calculations}
\item \textbf{Light-induced symmetry breaking and hopping}
\end{enumerate}
\end{multicols}

\renewcommand{\figurename}{Fig.~S}

\clearpage

\baselineskip24pt %

\section*{1. trTDTS measurements}\vspace{3mm}

\subsection*{Equilibrium optical properties}\vspace{-2mm}

\begin{figure}[h]
    \centering
    \includegraphics[width=1\textwidth]{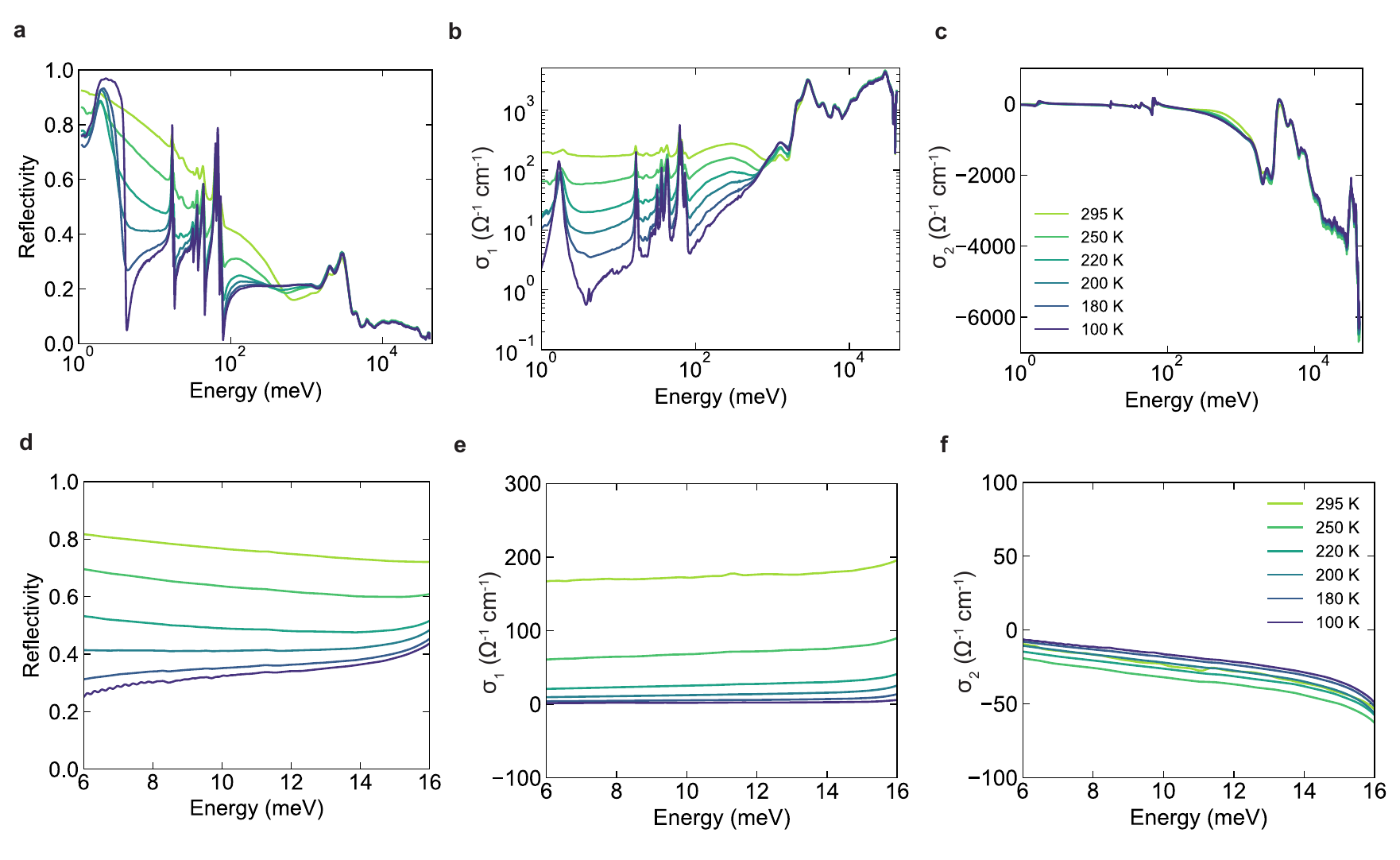}
    \caption{ {\bf Fig. ~S1. Equilibrium optical properties.} \textbf{a}, Broadband equilibrium reflectivity, and \textbf{b}, real ($\sigma_1$) and \textbf{c}, imaginary ($\sigma_2$) parts of the optical conductivity, as a function of temperature.  \textbf{d}, Broadband equilibrium reflectivity, and \textbf{e}, real ($\sigma_1$) and \textbf{f}, imaginary ($\sigma_2$) parts of the optical conductivity, as a function of temperature, within the frequency range of our time-resolved time-domain THz spectroscopy measurements.}
    \label{fig:opticalConductivity}
\end{figure}

In Fig. S\ref{fig:opticalConductivity}a, we present the temperature-dependent broadband reflectivity of Sr$_{14}$Cu$_{24}$O$_{41}$, previously reported in \cite{thorsmolle2012phonon}. Using Kramers-Kronig relations, we obtained the real ($\sigma_1$) and imaginary ($\sigma_2$) parts of the optical conductivity, shown in Fig. S\ref{fig:opticalConductivity}b and c, respectively. The spectra indicate an insulating state at 100 K, which becomes progressively more conducting with increasing temperature, as evidenced by the enhanced low-frequency spectral weight in $\sigma_1$. These observations indicate a gapped charge-ordered ground state with an onset below $T_\mathrm{CO}$ = 250 K, consistent with previous reports \cite{vuletic2003suppression, abbamonte2004crystallization, thorsmolle2012phonon}.

\begin{figure}[h]
    \centering
    \includegraphics[width=0.5\textwidth]{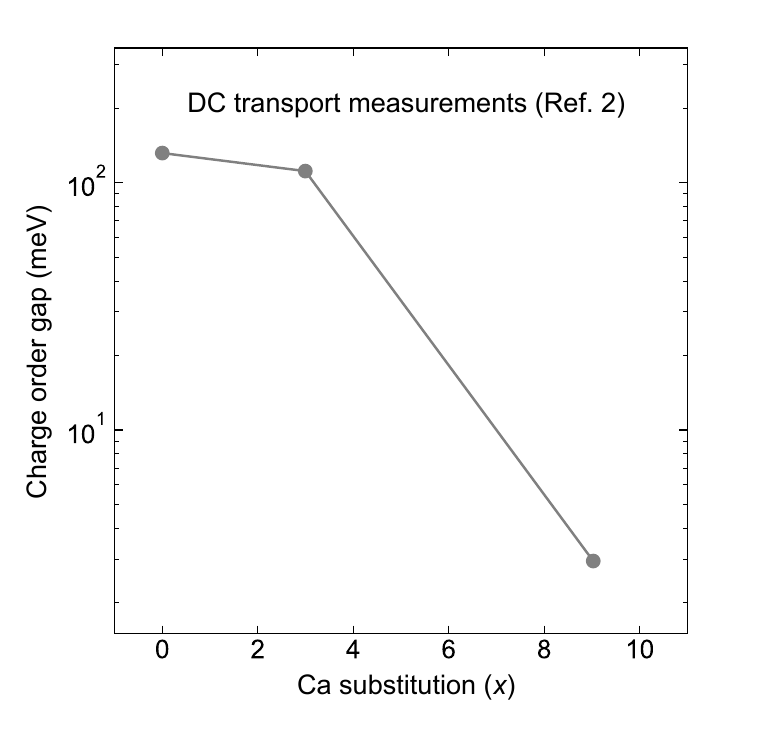}
    \caption{ {\bf Fig. ~S2. Charge order gap as a function of Ca substitution.} The charge order gap extracted from DC transport measurements, as a function of Ca substitution $x$ in Sr$_{14-x}$Ca$_x$Cu$_{24}$O$_{41}$, reproduced from \cite{vuletic2003suppression}.}
    \label{fig:cdwGapVsCa}
\end{figure}

By closely examining $\sigma_1$, we note that the charge order gap remains unchanged with temperature. This behavior is in contrast to $\sigma_1$ in the light-induced metastable phase (see Fig. 2d of the main text), which exhibits a suppressed charge order gap of the order of 10 meV. Notably, the charge order gap exhibits a similar suppression when Sr$_{14}$Cu$_{24}$O$_{41}$ is substituted with Ca, as reported in \cite{vuletic2003suppression} and reproduced in Fig. S\ref{fig:cdwGapVsCa}. The metastable state in our measurements exhibits a gap that is intermediate between that of the $x$ = 3 and $x$ = 9 compounds.

\clearpage

\subsection*{trTDTS: Fit functions and parameters}\vspace{-2mm}

We model the transient reflectivity in Fig. 2b of the main text by an exponential decay, with a functional form given by
\begin{equation}
\label{eq1}
R(t) = R_0 + \frac{1}{2}\left[1 + \mathrm{erf}\left(\frac{t - t_0}{\tau_0}\right)\right]Ae^{-\left(\frac{t-t_0}{\tau_1}\right)},
\end{equation}
where $R_0$ is the reflectivity at equilibrium, $t_0$ is an arbitrary temporal offset, and $\tau_1$ is an exponential decay time constant. The error function term describes the initial enhancement of reflectivity upon pump excitation, characterized by a time constant $\tau_0$. The fit parameters are shown in Table \ref{tab1}.

\begin{table}[h]
\caption{\textbf{Table 1}. Fit parameters in Fig. 1c.}\label{tab1}%
\begin{tabular}{@{}llllll@{}}
\toprule
   & $R_0$ & $t_0$ (ps) & $\tau_0$ (ps) & $\tau_1$ (ns) & $A$ \\
\midrule
$\omega$ = 15 meV & 0.372(4)    & 0.01(2)   & 0.57(5)  & 2.6(5) & 0.383(4)  \\
\botrule
\end{tabular}

\end{table}

\clearpage

\section*{2. trXRD measurements}\vspace{3mm}

\subsection*{Fit functions and parameters}\vspace{-2mm}

\begin{figure}[h]
    \centering
    \includegraphics[width=1\textwidth]{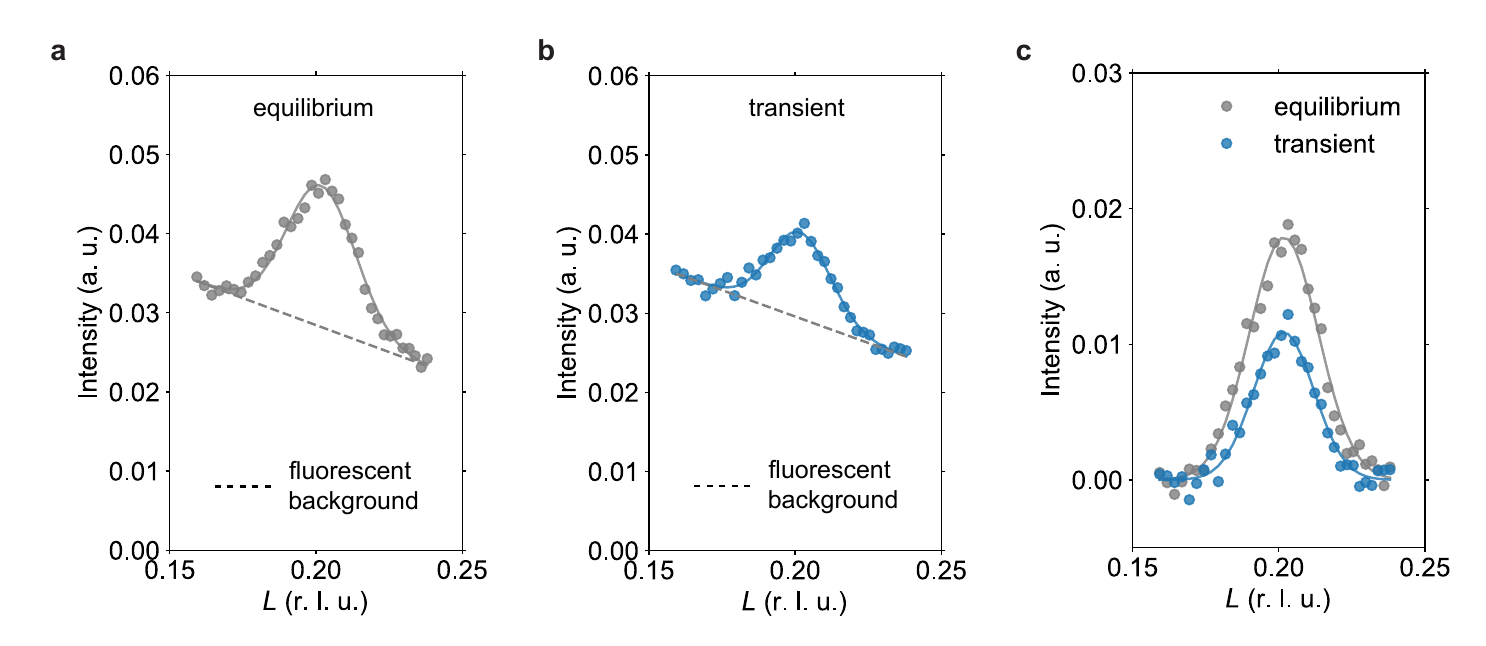}
    \caption{ {\bf Fig. ~S3. Background subtraction in trXRD data.} \textbf{a-b}, Equilibrium (a) and transient (b) time-resolved x-ray diffraction at the charge order peak, measured resonant with the O $K$-edge. The fluorescent background is denoted by dashed grey lines. \textbf{c}, Diffraction data from panels a and b with the fluorescent background subtracted. Solid lines denote fits.}
    \label{fig:trxrdBg}
\end{figure}

The data in Fig. 3b of the main text are obtained by fitting and subtracting a fluorescent background. We detail this procedure here. In Fig. S\ref{fig:trxrdBg}, we show the equilibrium and transient reciprocal space scans of the charge order diffraction peak in our time-resolved x-ray diffraction (trXRD) measurements. The charge order peak occurs over a large fluorescent background. We model these data by the sum of a linear function and a Gaussian, which describe the fluorescent background and charge order diffraction peak, respectively. The functional form is
\begin{equation}
\label{eq2}
I(L) = G(L; A, L_0, \sigma) + aL + b.
\end{equation}
Here, $I$ is the intensity, and $L$ is the momentum transfer along $c$. $G(L; A, L_0, \sigma)$ is a Gaussian function
\begin{equation}
G(L; A, L_0, \sigma) = \frac{A}{\sigma\sqrt{2\pi}}e^{{-\frac{(L - L_0)^2}{2\sigma^2}}},
\end{equation}
where $A$ is the amplitude, $L_0$ is the center, and $\sigma$ is the width. The fit parameters are shown in Table \ref{tab2}. We subtract the fluorescent background shown in Fig. S\ref{fig:trxrdBg}a-b to obtain the background-subtracted data in Fig. S\ref{fig:trxrdBg}c and Fig. 3b of the main text.

\begin{table}[h]
\caption{\textbf{Table 2}. Fit parameters in Fig. 3b and Fig. S\ref{fig:trxrdBg}.}\label{tab2}%
\begin{tabular}{@{}llllll@{}}
\toprule
   & $A$ (a. u.) & $L_0$ (r. l. u.) & $\sigma$ (r. l .u) & $a$ ($c_\mathrm{L}$) & $b$ (a. u.)\\
\midrule
Equilibrium    & 5.3(2)   & 0.2018(4)  & 0.0118(4)  & -0.136(8) & 0.056(1) \\
Transient    & 2.8(1)   & 0.2017(4)  & 0.0103(5)  & -0.133(6) & 0.056(1) \\
\botrule
\end{tabular}

\end{table}

Next, we describe the fitting of the delay traces in Fig. 3c of the main text. We model the time delay traces by an error function given by
\begin{equation}
\label{eq3}
I(t) = 1 - \frac{1}{2}A\left[1 + \mathrm{erf}\left(\frac{t - t_0}{\tau_0}\right)\right],
\end{equation}
where $\tau_0$ describes the timescale of pump-induced charge order suppression convolved with the pump-probe autocorrelation, $A$ is the magnitude of charge order suppression, and $t_0$ is an arbitrary temporal offset. Fitting the delay trace in Fig. 3c, we evaluate the time constant $\tau_0$ of charge order suppression to be 0.19(3) ps. The fit parameters are listed in Table \ref{tab3}.

\begin{table}[h]
\caption{\textbf{Table 3}. Fit parameters in Fig. 3c.}\label{tab3}%
\begin{tabular}{@{}llll@{}}
\toprule
   & $A$ (norm.) & $t_0$ (ps) & $\tau_0$ (ps)\\
\midrule
$E_\mathrm{pump} \parallel a$ & 0.421(7) & 0.01(4) & 0.19(3)  \\
\botrule
\end{tabular}

\end{table}

\clearpage

\subsection*{Fluence dependence of light-induced charge order suppression}\vspace{-2mm}

\begin{figure}[h]
    \centering
    \includegraphics[width=0.8\textwidth]{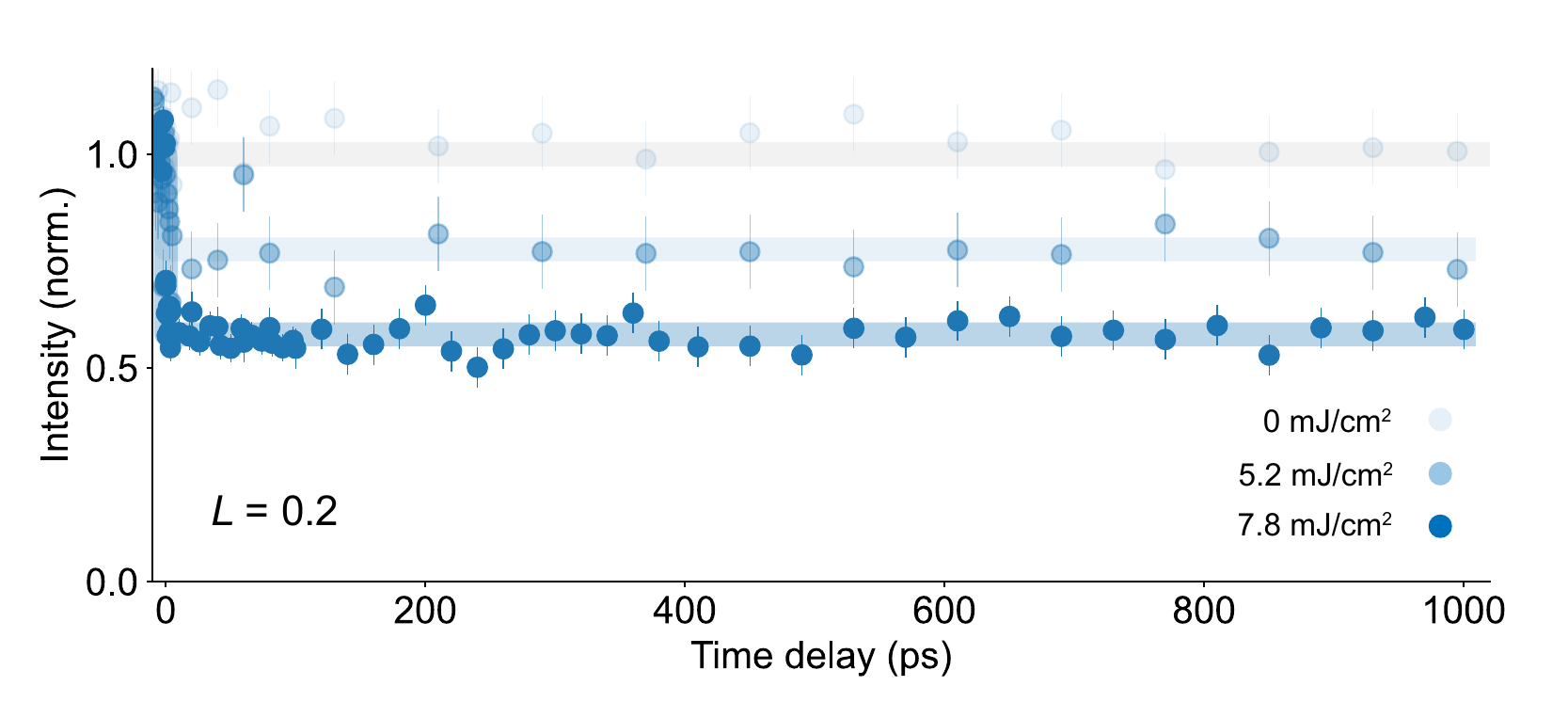}
    \caption{ {\bf Fig. ~S4. Fluence dependence of charge order suppression.} The light-induced charge order suppression is metastable up to nanosecond timescales, independent of the incident fluence. Error bars are the standard deviation of the signal at negative time delays. Solid lines denote fits.}
    \label{fig:trxrdFluence}
\end{figure}

We measure the charge order response as a function of pump fluence, with $E_\mathrm{pump} \parallel a$, as shown in Fig. S\ref{fig:trxrdFluence}. We find that the light-induced charge order suppression exhibits no recovery up to nanosecond timescales, independent of the incident fluence. 

We fix $\tau_0$ to the same value as in Table 3 and fit the data in Fig. S\ref{fig:trxrdFluence}. The fit results are shown in Table \ref{tab4}.

\begin{table}[h]
\caption{\textbf{Table 4}. Fit parameters in Fig. S4.}\label{tab4}%
\begin{tabular}{@{}llll@{}}
\toprule
   & $A$ (norm.) & $t_0$ (ps) & $\tau_0$ (ps)\\
\midrule
5.2 mJ/cm$^2$ & 0.22(1) & 0.1(2) & 0.19 (fixed)  \\
7.8 mJ/cm$^2$ & 0.421(7) & 0.01(4) & 0.19 (fixed)  \\
\botrule
\end{tabular}

\end{table}

\clearpage

\section*{3. trXAS measurements at Cu $L$-edge}\vspace{3mm}

\subsection*{Equilibrium XAS as a function of Ca substitution}\vspace{-2mm}

\begin{figure}[h]
    \centering
    \includegraphics[width=0.45\textwidth]{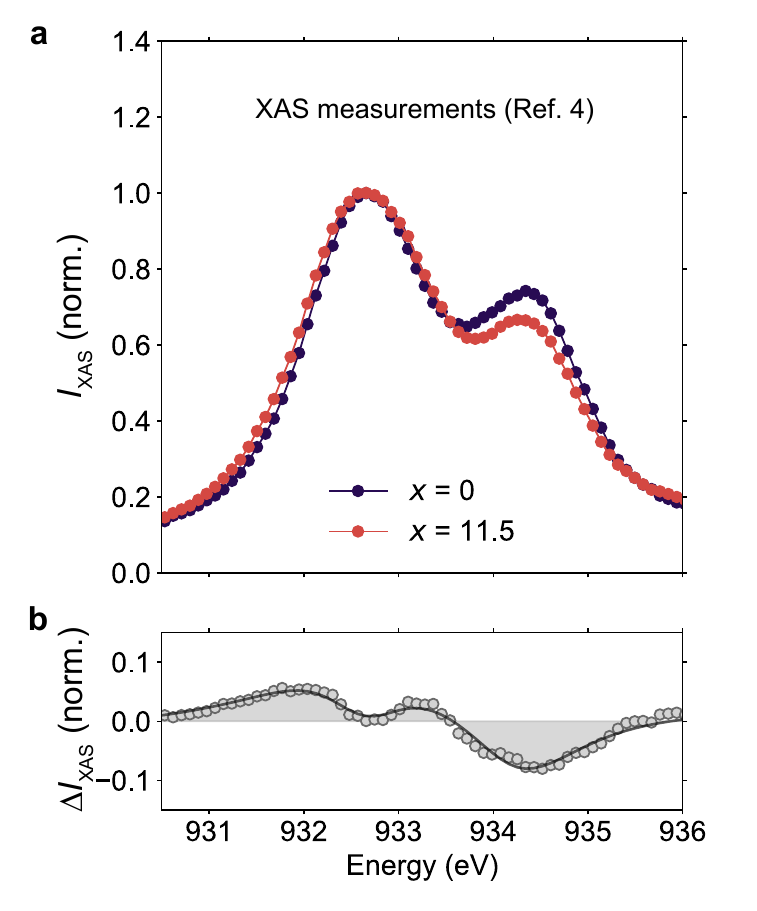}
    \caption{ {\bf Fig. ~S5. Cu $L$-edge XAS as a function of Ca substitution.} The static Cu $L$-edge x-ray absorption spectra (XAS) as a function of Ca substitution $x$ in Sr$_{14-x}$Ca$_x$Cu$_{24}$O$_{41}$, extracted from \cite{nucker2000hole}. The top panel shows the raw spectra and the bottom panel shows the difference between the $x$ = 0 and $x$ = 11.5 spectra. Solid lines are fits, as described in the text.}
    \label{fig:cuXasCaDoping}
\end{figure}

The primary effect of Ca substitution in Sr$_{14-x}$Ca$_{x}$Cu$_{24}$O$_{41}$ is the transfer of holes from chain to ladder subunits. This hole transfer presents a characteristic signature in the Cu $L$-edge x-ray absorption spectra (XAS), as shown in Fig. 4d of the main text. Here, we provide a detailed account of the Cu $L$-edge XAS as a function of Ca substitution at equilibrium. In Fig. S\ref{fig:cuXasCaDoping}, we reproduce the XAS measured on the $x$ = 0 and $x$ = 11.5 compounds, previously reported in \cite{nucker2000hole}. The two spectral features in Fig. S\ref{fig:cuXasCaDoping} can be distinguished based on the local bonding environment, as outlined in the main text. The peak centered at 932.7 eV consists of contributions primarily from the corner-shared ladder, while the peak centered at 934.4 eV is from the edge-shared chain \cite{huang2013determination}. We model the two-peaked spectra by the sum of two Lorentzians and a linear function as
\begin{equation}
\label{eq4}
I(\omega) = L(\omega; A_1, \omega_{01}, \sigma_1) + L(\omega;, A_2, \omega_{02}, \sigma_2) + a\omega + b,
\end{equation}
where the Lorentzians $L(\omega; A_i, \omega_{0i}, \sigma_i)$ describe ladder (i = 1) and chain (i = 2) peaks. The Lorentzian function is given by
\begin{equation}
L(\omega; A, \omega_{0}, \sigma) = \frac{A}{\pi}\left[\frac{\sigma}{(\omega - \omega_0)^2 + \sigma^2}\right],
\end{equation}
where $A$, $\omega_0$, and $\sigma$ are the amplitude, center energy, and linewidth parameters. The fit parameters are shown in Table \ref{tab5}.

\begin{table}[h]
\caption{\textbf{Table 5}. Fit parameters in Fig. S\ref{fig:cuXasCaDoping} and Fig. 4d.}\label{tab5}%
\begin{tabular}{@{}lllllll@{}}
\toprule
   & $A_1$ (norm.) & $\omega_{01}$ (eV) & $\sigma_1$ (eV) & $A_2$ (norm.)  & $\omega_{02}$ (eV)  & $\sigma_2$ (eV)\\
\midrule
$x = 0$ & 2.34(4) & 932.650(5) & 0.81(1) & 1.19(3)  & 934.398(8)  & 0.69(2) \\
$x = 11.5$ & 2.70(4) & 932.650(5) & 0.90(1) & 0.88(3)  & 934.42(1) & 0.65(2) \\

\botrule
\end{tabular}
\end{table}

We observe a reshaping of the XAS spectrum due to chain-to-ladder hole transfer in the form of a suppression of the amplitude $A_2$ of the chain peak and an enhancement of the amplitude $A_1$ of the ladder peak as a function of Ca substitution $x$. 

\begin{figure}[h]
    \centering
    \includegraphics[width=0.8\textwidth]{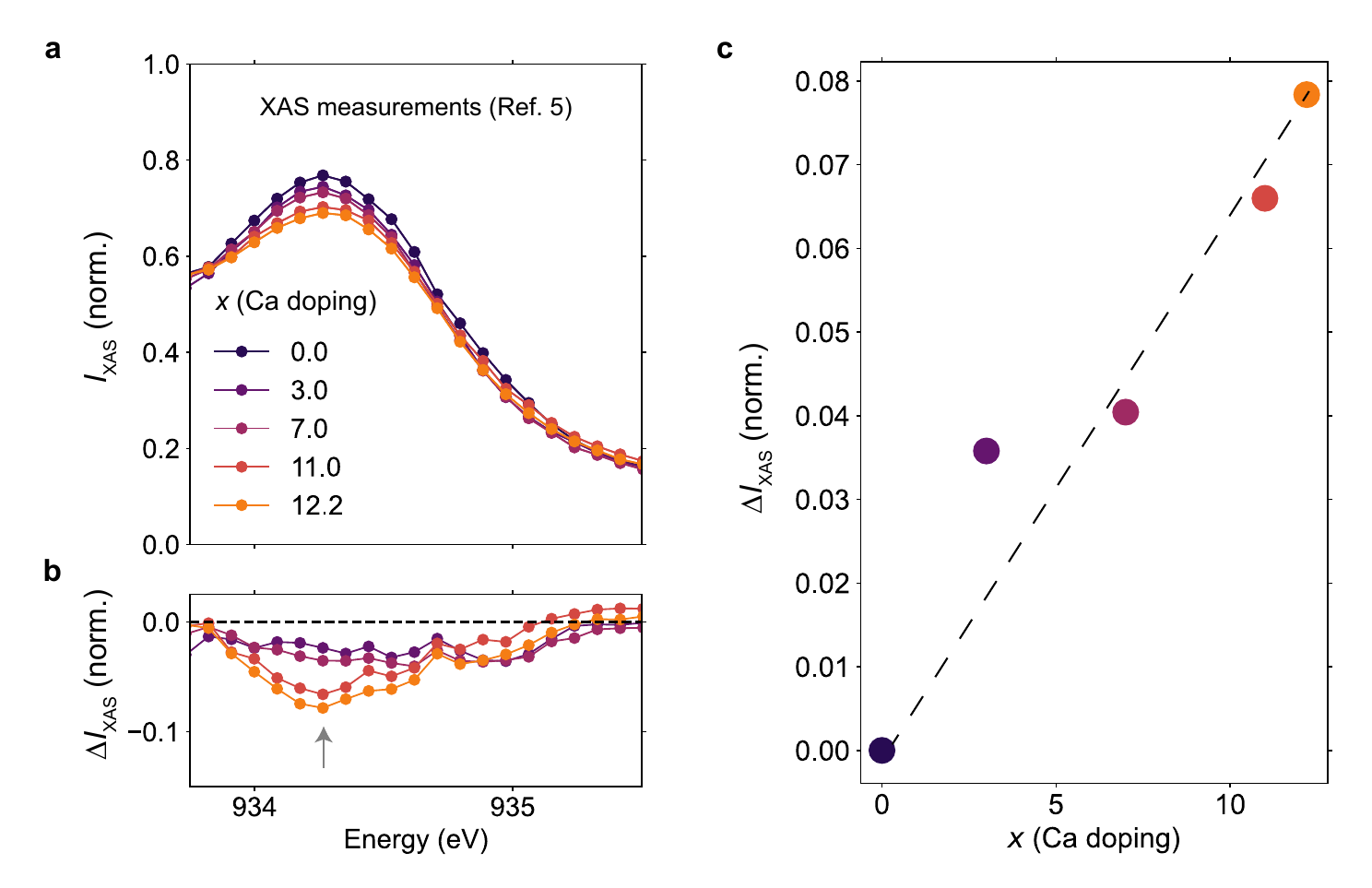}
    \caption{ {\bf Fig. ~S6. Linearity of XAS response to Ca substitution.} \textbf{a}, The static Cu $L$-edge x-ray absorption spectra (XAS) as a function of Ca substitution $x$ in Sr$_{14-x}$Ca$_x$Cu$_{24}$O$_{41}$, extracted from \cite{huang2013determination}. \textbf{b}, The difference spectra with respect to the $x$ = 0 spectrum in panel (a). \textbf{c}, The difference intensity at 934.4 eV, denoted by the arrow in panel (b), is plotted as a function of Ca substitution $x$. The dashed line is a guide to the eye.}
    \label{fig:cuXasCaDopingLinearity}
\end{figure}

We further show that the spectral reshaping is linear with respect to the Ca substitution $x$. In Fig. S\ref{fig:cuXasCaDopingLinearity}, we plot the XAS spectra as a function of $x$, previously reported in \cite{huang2013determination}, focusing on the well-defined suppression in the high-energy peak associated with the chain. We observe that the suppression is monotonic and linear in $x$, as shown in Fig. S\ref{fig:cuXasCaDopingLinearity}c. Given that the magnitude of chain-to-ladder hole transfer varies linearly with $x$ \cite{osafune1997optical, nucker2000hole, huang2013determination, Tseng2022crossover}, the magnitude of the spectral reshaping is linear with respect to $p$.

\begin{figure}[h]
    \centering
    \includegraphics[width=1\textwidth]{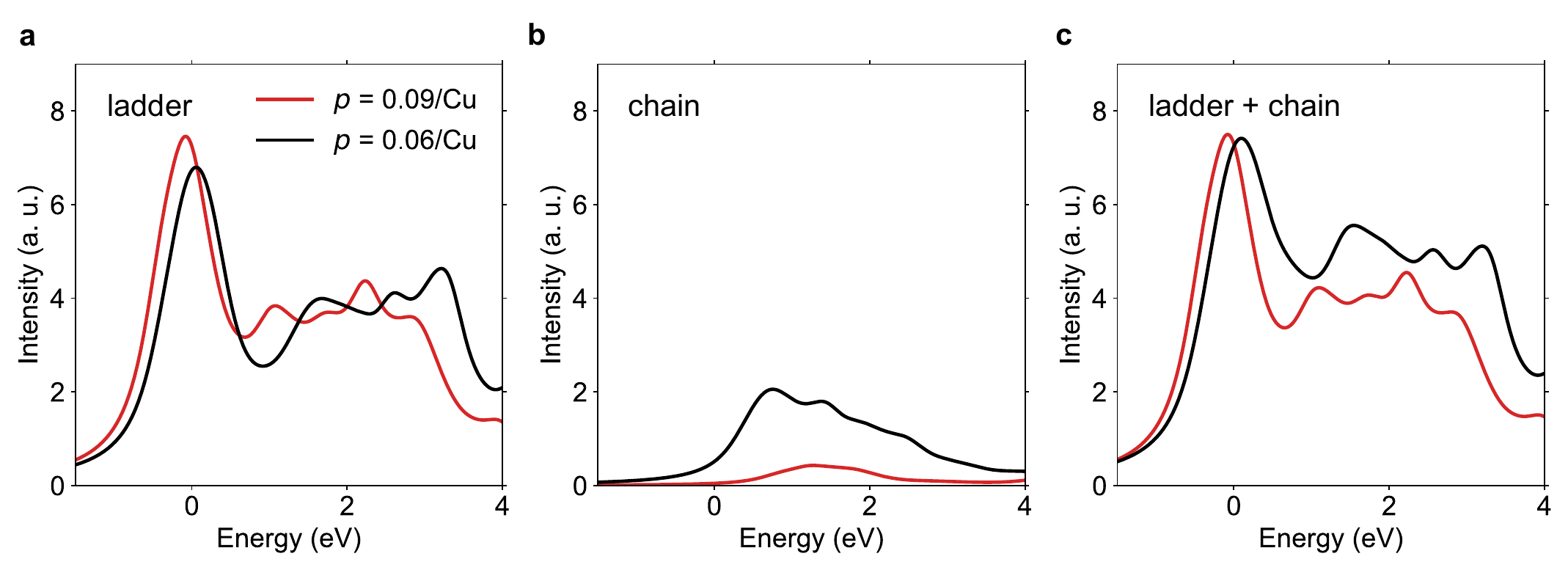}
    \caption{ {\bf Fig. ~S7. Spectral decomposition of Cu L edge XAS calculated by DFT.} Density functional theory (DFT) calculation of the Cu $L$-edge x-ray absorption spectra (XAS) of \textbf{a}, the ladder subunit and \textbf{b}, the chain subunit, for ladder hole densities $p$ = 0.06/Cu$_\mathrm{L}$ and 0.09/Cu$_\mathrm{L}$. \textbf{c}, The sum of the spectra in panels (a) and (b).}
    \label{fig:dftXas}
\end{figure}

Finally, we confirm our interpretation of the spectral reshaping using density functional theory (DFT) calculations. We calculated the Cu $L$-edge XAS spectrum for isolated ladder and chain subunits, as outlined in the Methods. We first examine the spectra corresponding to the equilibrium state, where the ladder hole density is $p$ = 0.06/Cu$_\mathrm{L}$. The ladder XAS spectrum shown in Fig. S\ref{fig:dftXas}a features a sharp peak and a continuum at higher energies. The chain XAS spectrum in Fig. S\ref{fig:dftXas}b is somewhat broader and much weaker than the ladder one. Next, we examine the calculated spectra corresponding to the light-induced metastable state, with ladder hole density $p$ = 0.09/Cu$_\mathrm{L}$. The sharp peak in the ladder XAS signal exhibits an increase in intensity and a small red shift, whereas the intensity of the chain XAS peak is strongly suppressed. These spectral changes are consistent with the addition of Cu $3d$ holes on the ladder and the reduction of of Cu $3d$ holes on the chain. These results are in agreement with our assignment of the experimental peaks and their reshaping upon chain-to-ladder hole transfer. Finally, we note that since our calculation consists of isolated chains and ladders with different total charges (and the ensuing background charge compensation), the relative energy offset between the chain and ladder spectra cannot be determined theoretically. To overcome this limitation, we estimate the appropriate energy offset from the available experimental data (Fig. S\ref{fig:cuXasCaDoping}). Our XAS data and previous XAS measurements at the Cu $L$ edge in the presence of Ca dopant ions define the energy positions of the ladder and chain peaks, respectively, with an energy separation of 1.6 eV. We apply the same offset to our simulated equilibrium and pumped spectra. Summing the ladder and chain spectra (Fig. S\ref{fig:dftXas}c), we find that the calculated XAS are consistent with the experimental XAS at equilibrium, reproducing both the suppression of the high-energy peak associated with the chains and the enhancement of the low-energy shoulder of the peak associated with the ladders.
\clearpage

\subsection*{trXAS measurements}\vspace{-2mm}

We model the trXAS spectra in Fig. 3a-b of the main text using the same functional form as in Eq. 5 and 6. The fit parameters are shown in Table \ref{tab6}.

\begin{table}[h]
\caption{\textbf{Table 6}. Fit parameters in Fig. 3a-b}\label{tab6}
\begin{tabular}{@{}lllllll@{}}
\toprule
   & $A_1$ (norm.) & $\omega_{01}$ (eV) & $\sigma_1$ (eV) & $A_2$ (norm.)  & $\omega_{02}$ (eV)  & $\sigma_2$ (eV)\\
\midrule
Equilibrium & 1.88(3) & 932.611(5) & 0.73(1) & 1.18(3)  & 934.333(8)  & 0.74(2) \\
Transient & 1.95(3) & 932.600(5) & 0.75(1) & 1.26(4)  & 934.32(1) & 0.83(2) \\

\botrule
\end{tabular}
\end{table}

\begin{figure}[h]
    \centering
    \includegraphics[width=1\textwidth]{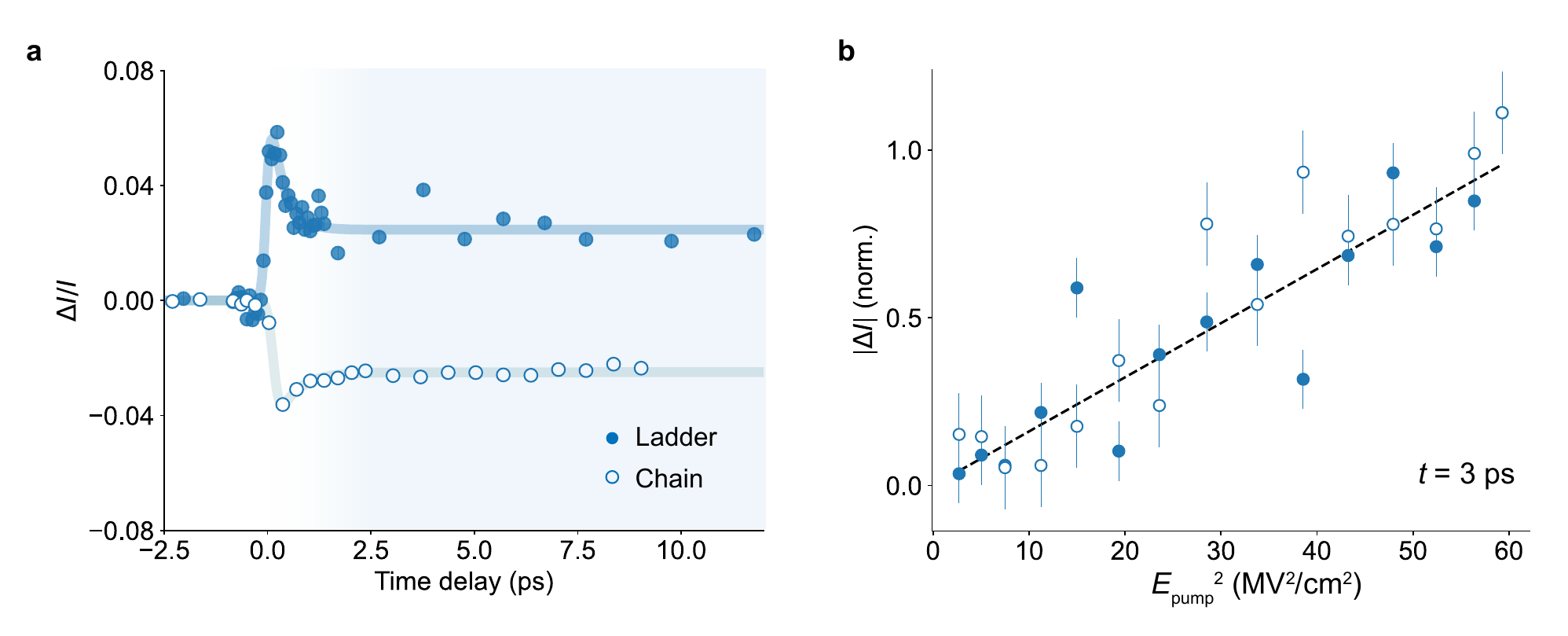}
    \caption{ {\bf Fig. ~S8. Temporal evolution and pump field dependence of trXAS spectra.} \textbf{a}, Traces of the temporal evolution at fixed energies, 932.0 eV (labeled `ladder') and 934.4 eV (labeled `chain'). Fits are denoted by solid lines. \textbf{b}, The normalized response as a function of the pump field at a time delay of 3 ps. The dashed line is a guide to the eye.}
    \label{fig:trxasDelayAndFluence}
\end{figure}

Next, we present the temporal evolution of the trXAS results described in Fig. 4 of the main text. The delay traces at energies fixed to 932.0 eV and 934.4 eV, corresponding to the ladder and chain peaks, respectively, are presented in Fig. S\ref{fig:trxasDelayAndFluence}a. The data show that the light-induced reshaping of the trXAS spectra has a short-lived initial response, followed by a long-lived metastable response. We model the time dependence using the following functional form:

\begin{equation}
I(t) = \frac{1}{2}\left[1 + \mathrm{erf}\left(\frac{t - t_0}{\tau_0}\right)\right]\left[B_0 + B_1e^\frac{(t-t_0)}{\tau_1}\right],
\end{equation}

where $\tau_0$ describes the timescale of the pump-induced response convolved with the pump-probe autocorrelation, $B_0$ is the magnitude of the metastable response, $B_1$ and $\tau_1$ are the amplitude and decay time constant of the fast response, respectively, and $t_0$ is an aribitrary temporal offset. Fitting the delay traces in Fig. S\ref{fig:trxasDelayAndFluence}a, we obtain the fit parameters listed in Table \ref{tab7}.

\begin{table}[h]
\caption{\textbf{Table 7}. Fit parameters in Fig. S8a}\label{tab7}
\begin{tabular}{@{}llllll@{}}
\toprule
   & $t_0$ (ps) & $\tau_0$ (ps) & $B_0$ (norm.) & $B_1$ (norm.) & $\tau_1$ (ps)\\
\midrule
Ladder & 0.02(2) & 0.13(3) & 0.123(6) & 0.267(4) & 0.33(8) \\
Chain & 0.1(1) & 0.2(1) & 0.125(2) & 0.10(3)  & 0.50(1) \\

\botrule
\end{tabular}
\end{table}

Finally, we present the field dependence of the light-induced reshaping of the trXAS spectra. In Fig. S\ref{fig:trxasDelayAndFluence}b, we plot the normalized response as a function of the square of the pump field, with the energies fixed at 932.0 eV (ladder) and 934.4 eV (chain). The response scales linearly with the square of the pump field, as indicated by the dashed line, consistent with the light-induced symmetry-breaking mechanism outlined in the main text and SI Section 7.

\clearpage

\section*{4. trXAS measurements at O $K$-edge}\label{sec4}

\begin{figure}[h]
    \centering
    \includegraphics[width=1\textwidth]{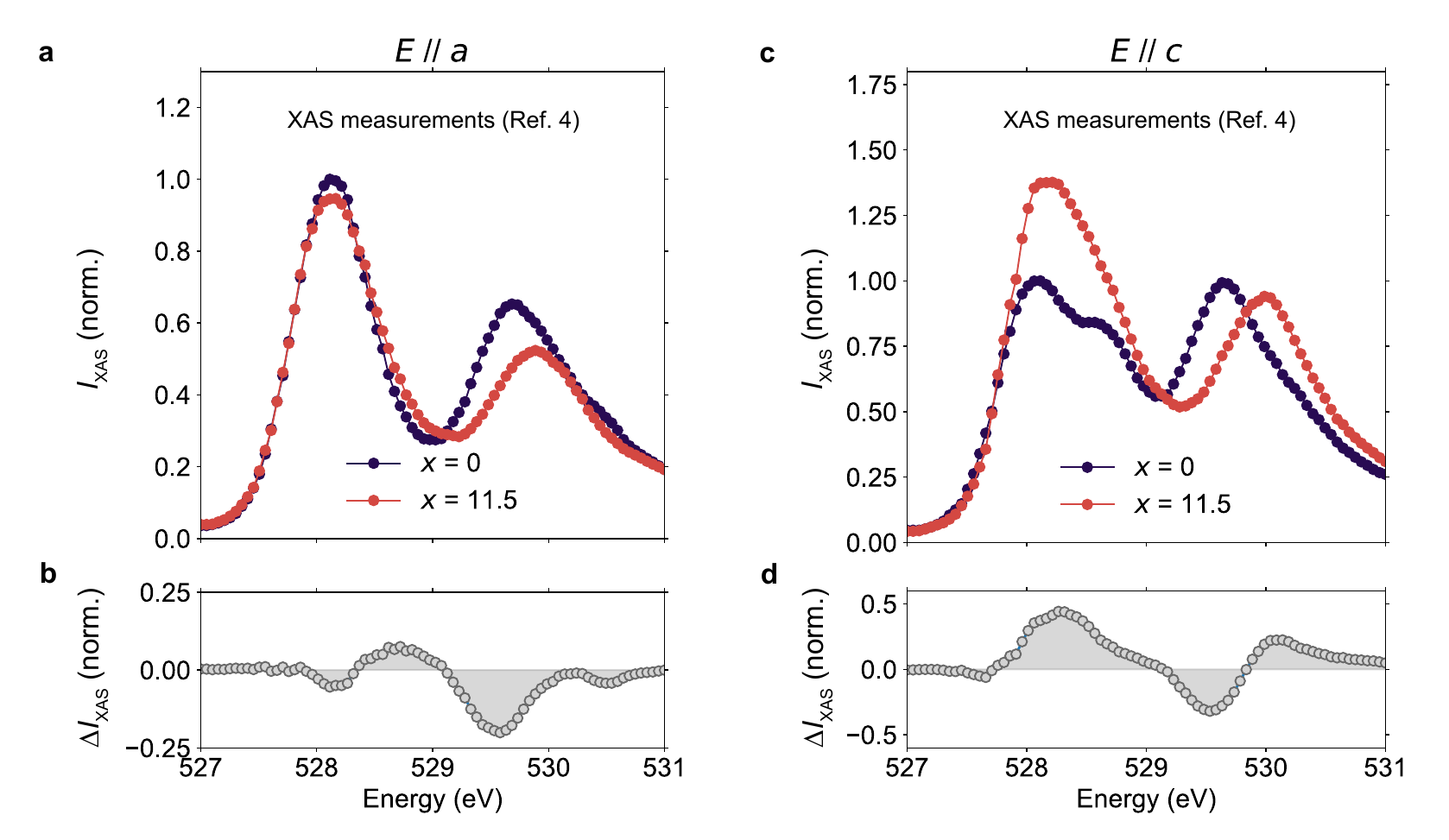}
    \caption{ {\bf Fig. ~S9. O $K$-edge XAS as a function of Ca substitution.} \textbf{a}, The static O $K$-edge x-ray absorption spectra (XAS) as a function of Ca substitution $x$ in Sr$_{14-x}$Ca$_x$Cu$_{24}$O$_{41}$ for x-rays polarized along $a$, extracted from \cite{nucker2000hole}. \textbf{b}, The difference between the $x$ = 0 and $x$ = 11.5 spectra shown in panel a. \textbf{c-d}, Same as panels a-b, for x-rays polarized parallel to $c$. Lines are a guide to the eye.}
    \label{fig:oXasCaDoping}
\end{figure}

We first examine the equilibrium O $K$-edge x-ray absorption spectra (XAS) as a function of Ca substitution for $x$ = 0 and $x$ = 11.5 (see Fig. S\ref{fig:oXasCaDoping}), which we reproduce here from Ref. \cite{nucker2000hole}. The spectra consist of two primary features: a peak centered at 528.2 eV, corresponding to the doped holes, also referred to as the Zhang-Rice singlet (ZRS) peak, and a peak centered at 529.5 eV, corresponding to the upper Hubbard band (UHB). In cuprates, an increase in the overall hole density is generally associated with a suppression of the UHB and an enhancement of the ZRS \cite{chen1991electronic}. While chain and ladder contributions to the O $K$-edge XAS cannot be unambiguously distinguished (unlike what is observed at the Cu $L$-edge) due to overlapping spectral features, previous studies have established clear qualitative trends associated with chain-to-ladder hole transfer \cite{nucker2000hole}, which we summarize below. 

With x-rays polarized parallel to $a$, the ZRS and UHB are both suppressed with increasing chain-to-ladder hole transfer. With x-rays polarized parallel to $c$, the ZRS features an additional peak centered at 528.7 eV that is attributed to holes in the ladder subunit. Hence, chain-to-ladder hole transfer in this case causes an enhancement of the ZRS together with a suppression of the UHB. Finally, we note that the UHB exhibits a significant blue shift with increasing Ca substitution, for both x-ray polarizations. The UHB of the ladder and chain subunits overlap in energy \cite{nucker2000hole}. If these changes were due to chain-to-ladder hole transfer, we would expect the ladder and chain peaks to shift in opposite directions, and hence lead to a substantial broadening of the UHB feature. In contrast, the change primarily manifests as an overall energy shift. This is more naturally explained in terms of the structural changes that occur when 82\% of the Sr atoms are replaced by Ca. For example, Ca substitution increases the displacement of the ladder oxygen along the $b$ axis \cite{deng2011structural}, which would modify the local bonding environment and hence influence the O $K$-edge XAS peak positions. Hence, we expect that light-driven chain-to-ladder hole transfer, which occurs without the structural distortions associated with Ca substitution, will manifest in the form of the spectral reshaping described above but without the additional blue shift of the UHB.

\begin{figure}[h]
    \centering
    \includegraphics[width=1\textwidth]{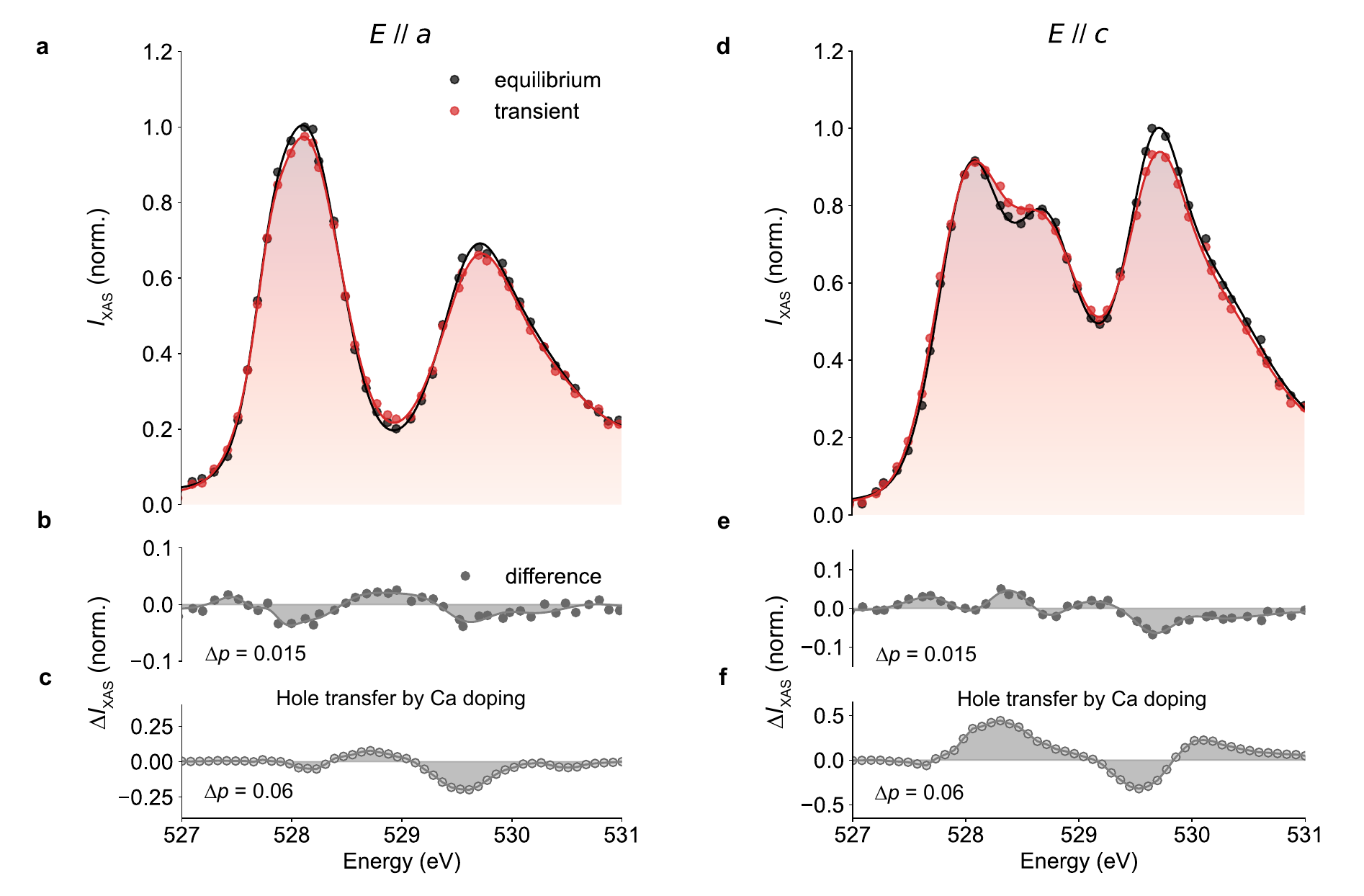}
    \caption{ {\bf Fig. ~S10. O $K$-edge trXAS} \textbf{a}, Equilibrium (black) and transient (red) O $K$-edge x-ray absorption spectra (XAS) at pump-probe delay $t$ = 3 ps, for x-rays polarized along $a$. \textbf{b}, Difference between equilibrium and transient XAS intensities.  \textbf{c}, Change in static XAS intensity due to chain-to-ladder hole transfer induced by Ca substitution in Sr$_{14-x}$Ca$_x$Cu$_{24}$O$_{41}$, derived from reference \cite{nucker2000hole} The data show the difference between XAS measured on the $x$ = 0 and $x$ = 11.5 compounds, which corresponds to a charge transfer of $\Delta p$ = 0.06 holes/Cu$_L$. \textbf{d-f} Same as panels a-c, for x-rays polarized parallel to $c$.}
    \label{fig:oTrxas}
\end{figure}

We show the results of our O $K$-edge trXAS measurements (pump polarization parallel to $c$ in all measurements) in Fig. S\ref{fig:oTrxas}. Note that the pump fluence of 3.9 mJ/cm$^2$ used in these measurements is half that used for the Cu $L$-edge trXAS measurements. The transient spectrum with x-ray polarization $E \parallel a$ exhibits a reshaping with both the ZRS and UHB suppressed relative to the equilibrium spectra, consistent with our expectations for chain-to-ladder hole transfer. The difference spectrum in S\ref{fig:oTrxas}b is in excellent agreement with the difference spectrum due to Ca substitution in Fig. S\ref{fig:oTrxas}c, rescaled by by a factor of 4. This corresponds to a hole transfer of 0.06/4 = 0.015 holes/Cu$_\mathrm{L}$. For x-ray polarization $E \parallel c$, shown in Fig. S\ref{fig:oTrxas}d, we observe a suppression of the UHB and an enhancement of the ZRS, consistent with a chain-to-ladder hole transfer without the structural distortions associated with Ca substitution. The lack of signatures of structural distortions in the light-induced hole transfer implies that light-induced doping is distinct from that due to Ca substitution at equilibrium. In summary, our O $K$-edge trXAS measurements corroborate the chain-to-ladder hole transfer detected and quantified by our Cu $L$-edge trXAS measurements described in the main text.

\clearpage

\section*{5. trRIXS measurements}\label{sec5}\vspace{3mm}

\subsection*{Triplon excitations}\vspace{-2mm}

\begin{figure}[h]
    \centering
    \includegraphics[width=0.8\textwidth]{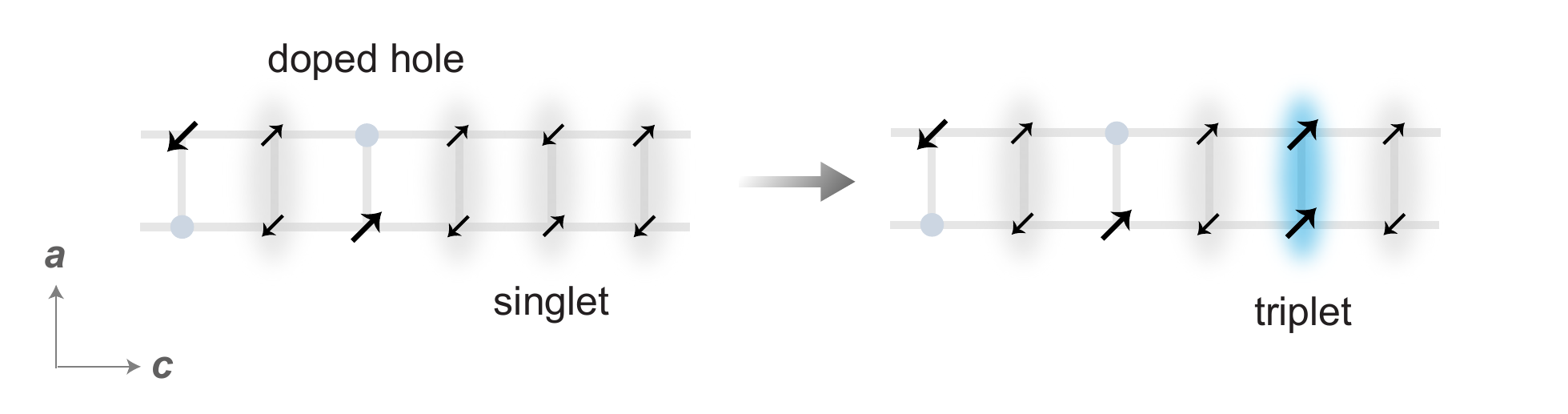}
    \caption{ {\bf Fig. ~S11. Triplon excitation} Schematic of dispersing singlet-to-triplet excitations called `triplons.' The black arrows represent spins, grey circles represent doped holes, grey clouds represent spin singlets, and the blue cloud represents a spin triplet excitation.}
    \label{fig:triplonSchematic}
\end{figure}

The ladder subunit of Sr$_{14}$Cu$_{24}$O$_{41}$ is composed of spin singlets. The fundamental magnetic excitations of this system are dispersive singlet-to-triplet excitations known as `triplons,' shown schematically in Fig. S\ref{fig:triplonSchematic}. The symmetry of the ladder geometry implies that sectors of even and odd triplon number do not mix in the RIXS spectra, due to different parity with respect to reflection about the center of the ladder legs. In particular, with $H$ = 0, as in our measurements, we only detect excitations with an even triplon number, which are dominated by the two-triplon continuum. We note that the inclusion of doped holes in the ladder disrupts the singlet background (see Fig. S\ref{fig:triplonSchematic}) and reduces the overall triplon scattering intensity, as discussed in the main text.

\subsection*{Subtraction of elastic peak}\vspace{-2mm}

The Cu $L$-edge RIXS spectra of Sr$_{14}$Cu$_{24}$O$_{41}$ consist of an elastic peak at zero energy loss, a phonon centered at 60 meV, triplons dispersing up to 400 meV for finite $L$, and $dd$ orbital excitations above 1.3 eV. Due to the combined energy resolution of our spectrometer, the elastic and phonon peaks cannot be distinguished from each other. We also note that for $\pi$ incident x-rays, as in our measurements, the phonon scattering intensity is much weaker than the elastic and spin-flip magnetic excitations. Given these two conditions, we assume that the convolution of the elastic and phonon peaks is approximately symmetric about zero energy loss. We fit the low-energy shoulder of this contribution to a Gaussian peak and subtract it from our RIXS spectra to isolate the magnetic scattering intensity below 1 eV. We assume that any residual contribution from the phonon is negligible compared to the magnetic excitations, especially away from the zone center ($L$ = 0).

\begin{figure}[h]
    \centering
    \includegraphics[width=0.85\textwidth]{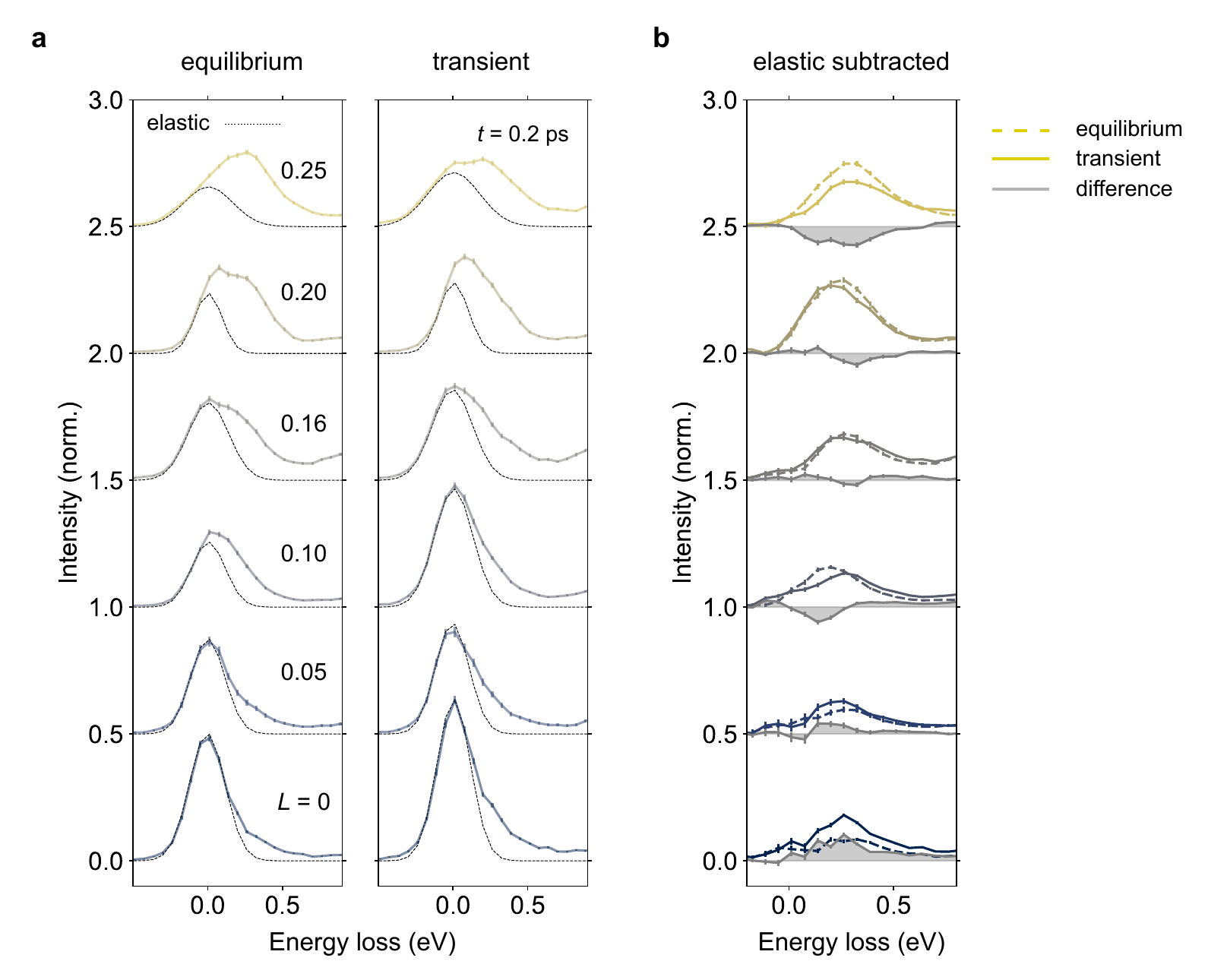}
    \caption{ {\bf Fig. ~S12. Raw trRIXS spectra at $t$ = 0.2 ps.} \textbf{a}, Equilibrium and transient time-resolved resonant inelastic x-ray scattering (trRIXS) spectra as a function of momentum $L$, at pump delay $t$ = 0.2 ps, normalized to the total intensity of $dd$ excitations. The elastic component is denoted by the dashed black line. \textbf{b}, trRIXS spectra with the elastic components subtracted. The difference between equilibrium and transient spectra is shown in grey.}
    \label{fig:trrixsT02}
\end{figure}

\begin{figure}[h]
    \centering
    \includegraphics[width=0.85\textwidth]{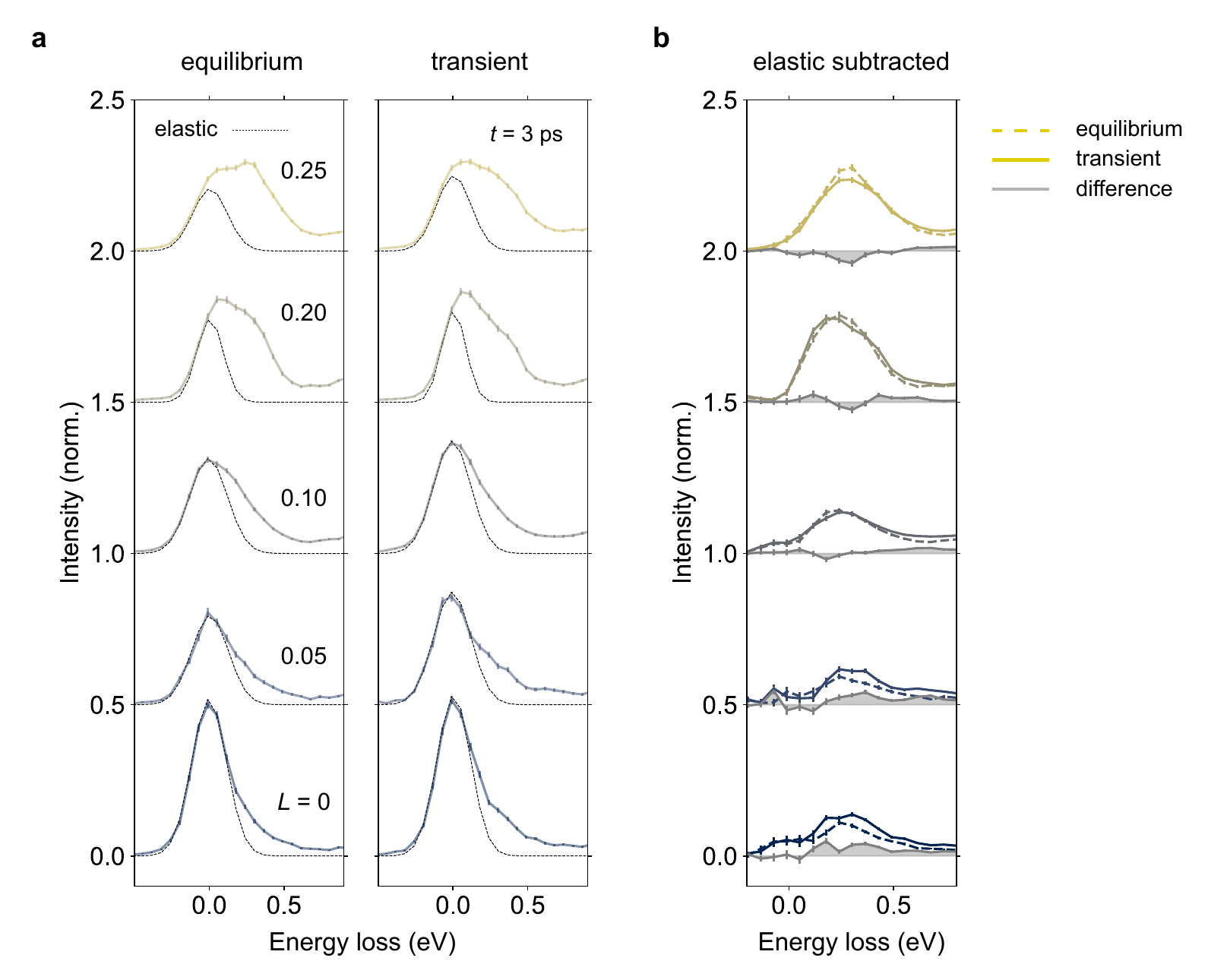}
    \caption{ {\bf Fig. ~S13. Raw trRIXS spectra at $t$ = 3 ps.} \textbf{a}, Equilibrium and transient time-resolved resonant inelastic x-ray scattering (trRIXS) spectra as a function of momentum $L$, at pump delay $t$ = 3 ps, normalized to the total intensity of $dd$ excitations. The elastic component is denoted by the dashed black line. \textbf{b}, trRIXS spectra with the elastic components subtracted. The difference between equilibrium and transient spectra is shown in grey.}
    \label{fig:trrixsT3}
\end{figure}

\begin{figure}[h]
    \centering
    \includegraphics[width=0.85\textwidth]{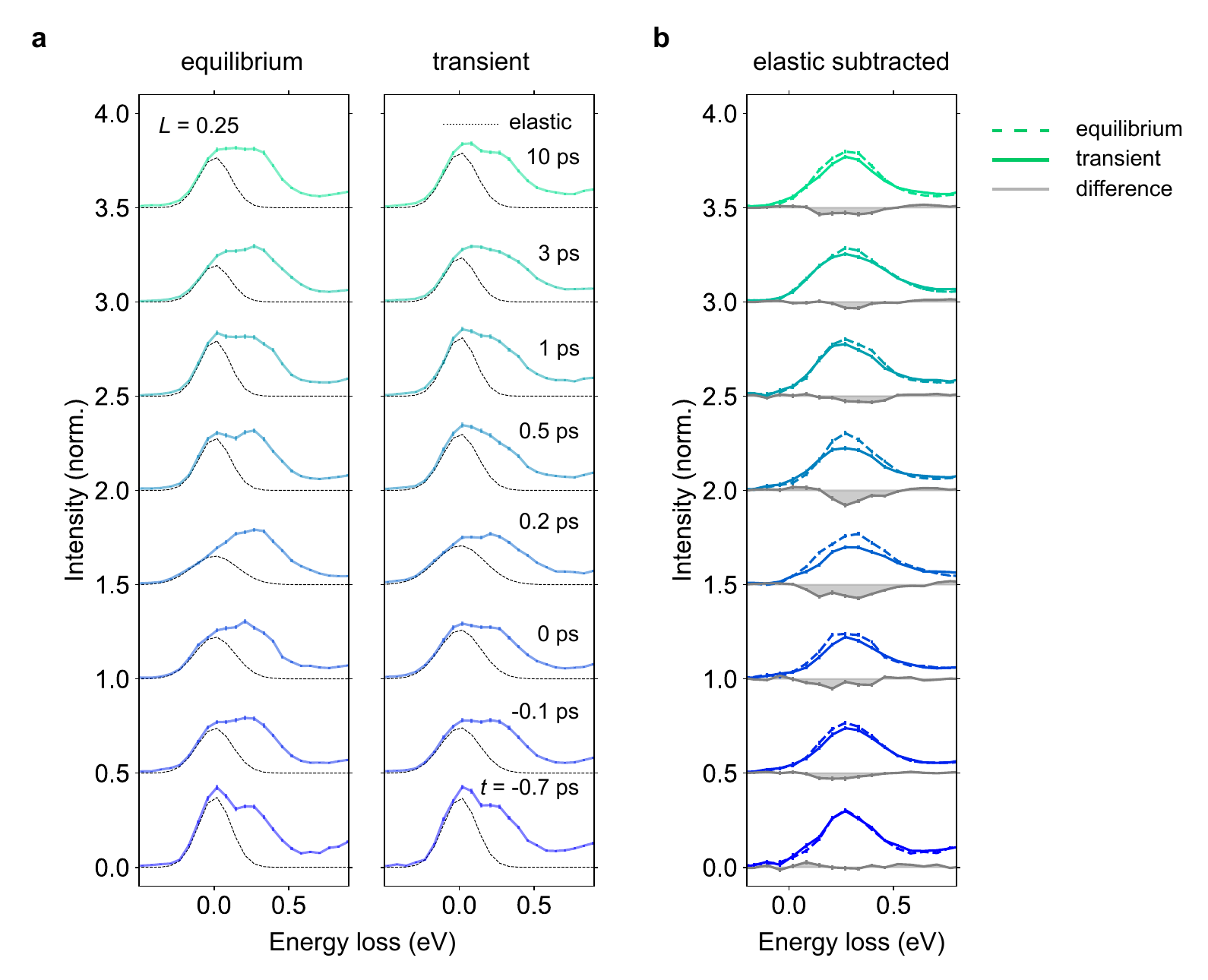}
    \caption{ {\bf Fig. ~S14. Raw trRIXS spectra at $L$ = 0.25.} \textbf{a}, Equilibrium and transient time-resolved resonant inelastic x-ray scattering (trRIXS) spectra as a function of pump delay $t$, at momentum $L$ = 0.25, normalized to the total intensity of $dd$ excitations. The elastic component is denoted by the dashed black line. \textbf{b}, trRIXS spectra with the elastic components subtracted. The difference between equilibrium and transient spectra is shown in grey. All spectra are normalized to the total intensity of $dd$ excitations.}
    \label{fig:trrixsQ025}
\end{figure}

We implement this elastic peak subtraction separately for the equilibrium and transient spectra at each measured momentum $L$ and pump delay $t$. In Fig. S\ref{fig:trrixsT02}, we show our results for trRIXS spectra measured as a function of $L$ with $t$ fixed to 0.2 ps. In Fig. S\ref{fig:trrixsT02}b, we show the spectra with the elastic peak subtracted. In Fig. S\ref{fig:trrixsT3}, we show the corresponding results for the $L$-dependent trRIXS spectra at $t$ = 3 ps, and in Fig. S\ref{fig:trrixsQ025}, the corresponding results for the $t$-dependent trRIXS spectra at $L$ = 0.25.

\clearpage

\subsection*{Extracting tr$S(q, \omega)$ from trRIXS spectra}\vspace{-2mm}
To extract the dynamical spin structure factor $S(q, \omega)$ from the RIXS spectra, we first subtract the non-magnetic components of the scattered signal, as outlined in the previous section. Next, we normalize the subtracted RIXS spectra by scaling them by a geometry-dependent factor to ensure the total integrated intensity of the orbital excitations matches the theoretical single-ion scattering cross-section calculated using exact diagonalization (ED) \cite{Wang2019EDRIXS}.  We do this by considering orbital excitations of a Cu$^{2+}$ ion in a square-planar crystal field, neglecting spin-flip and phonon contributions. In this section, $x$, $y$, and $z$ are parallel to the crystallographic $a$, $c$, and $b$ directions, respectively. We fix the crystal field parameters by matching the theoretical spectra to the experimental orbital excitation energies, and obtain $D_q$ = 0.164 eV, $D_s$ = 0.42 eV, and $D_t$ = 0.19 eV. In Fig. S\ref{fig:trrixsS(q,w)}a-b, we show the experimental and calculated RIXS spectra as a function of momentum $L$. The integrated intensity of orbital excitations is plotted as a function of the incident angle $\theta$ in Fig. S\ref{fig:trrixsS(q,w)}c. We multiply the experimental spectra in Fig. S\ref{fig:trrixsS(q,w)}a by a $\theta$-dependent factor to match this dependence within an overall scale factor. We refer to these spectra as the `normalized RIXS spectra.' 

\begin{figure}[h]
    \centering
    \includegraphics[width=0.9\textwidth]{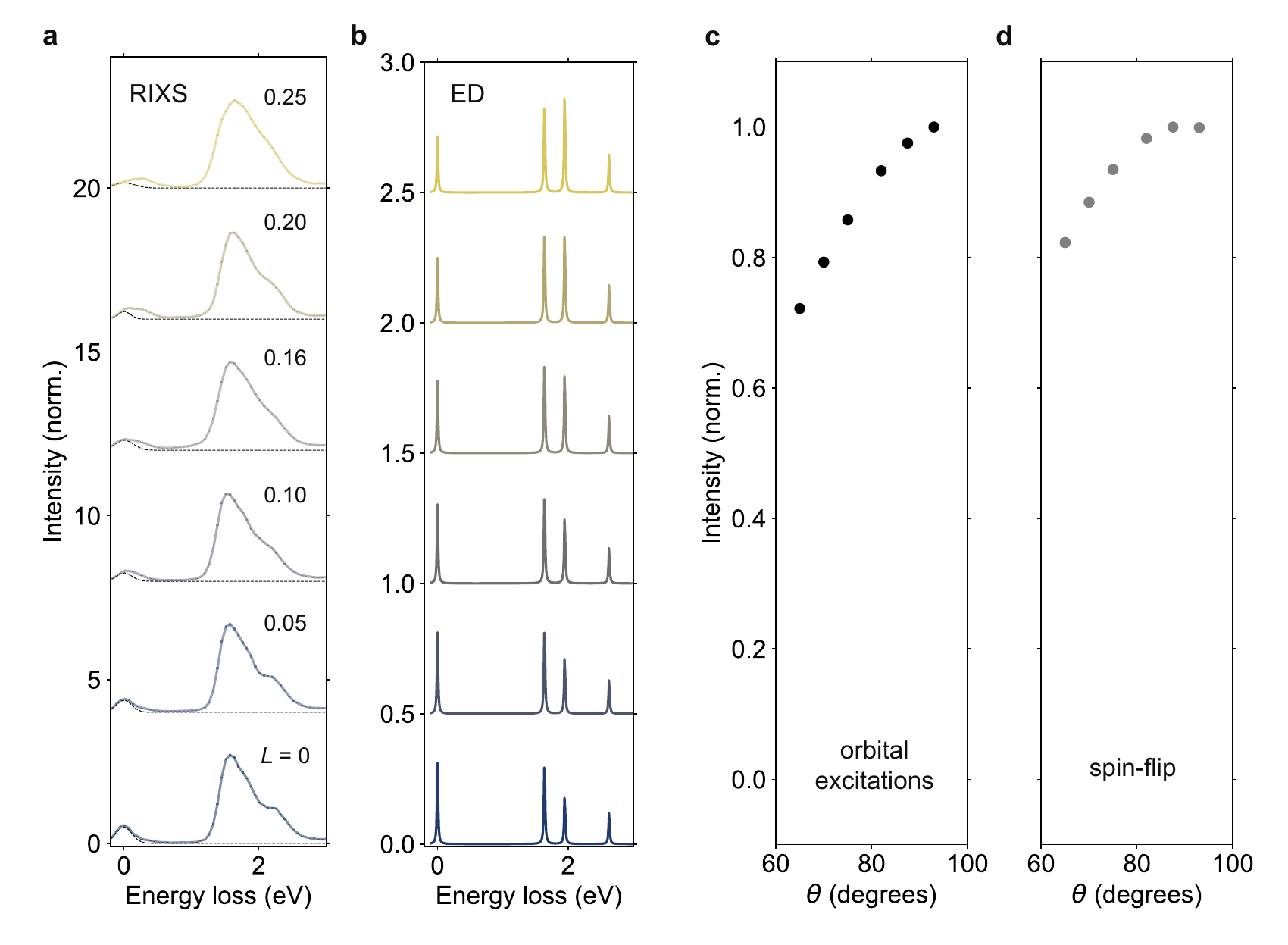}
    \caption{ {\bf Fig. ~S15. ED calculations to extract $S(q,\omega)$ from RIXS.} \textbf{a}, Equilibrium resonant inelastic x-ray scattering (RIXS) spectra as a function of momentum $L$, normalized to the total intensity of $dd$ excitations. \textbf{b}, Orbital excitation spectra calculated using exact diagonalization (ED) for $\pi$-polarized x-rays. \textbf{c-d}, Integrated orbital excitation scattering cross-section (c) and spin-flip cross-section (d) as a function of incident angle $\theta$.}
    \label{fig:trrixsS(q,w)}
\end{figure}

Finally, we extract $S(q, \omega)$ as follows. The RIXS magnetic excitation signal is proportional to $S(q, \omega)$ multiplied by the form factor of the single-ion spin-flip scattering cross-section, $R_\mathrm{spin}(\epsilon, \epsilon', \Omega_i)$, where $\Omega_i$ is the excitation energy and $\epsilon$ and $\epsilon'$ are the polarizations of incident and scattered photons \cite{ament2009theoretical, Robarts2021dynamical}
\begin{equation}
\label{eq6}
I_\mathrm{spin} \propto R_\mathrm{spin}(\epsilon, \epsilon', \Omega_i)\times S(q, \omega).
\end{equation}
Even though this formalism does not account for hole doping, its validity for doped systems within the energy range of magnetic excitations has been verified in Ref. \cite{jia2014persistent}. We calculate $R_\mathrm{spin}$ for a hole in a $d_{x^2-y^2}$ orbital with its spin oriented in-plane along [1, 0, 1], following the approach in previous publications \cite{ament2009theoretical, shen2022role}. The results are plotted in Fig. S\ref{fig:trrixsS(q,w)}d. We obtain $S(q,\omega)$ up to an overall scaling factor by dividing the normalized RIXS spectra by $R_\mathrm{spin}$. We apply these steps to both the equilibrium and transient RIXS spectra, subtracting them to obtain the time-resolved $S(q, \omega)$ spectra shown in Fig. 5 of the main text.

\subsection*{Temporal evolution of tr$S(q, \omega)$}\vspace{-2mm}

\begin{figure}[h]
    \centering
    \includegraphics[width=0.9\textwidth]{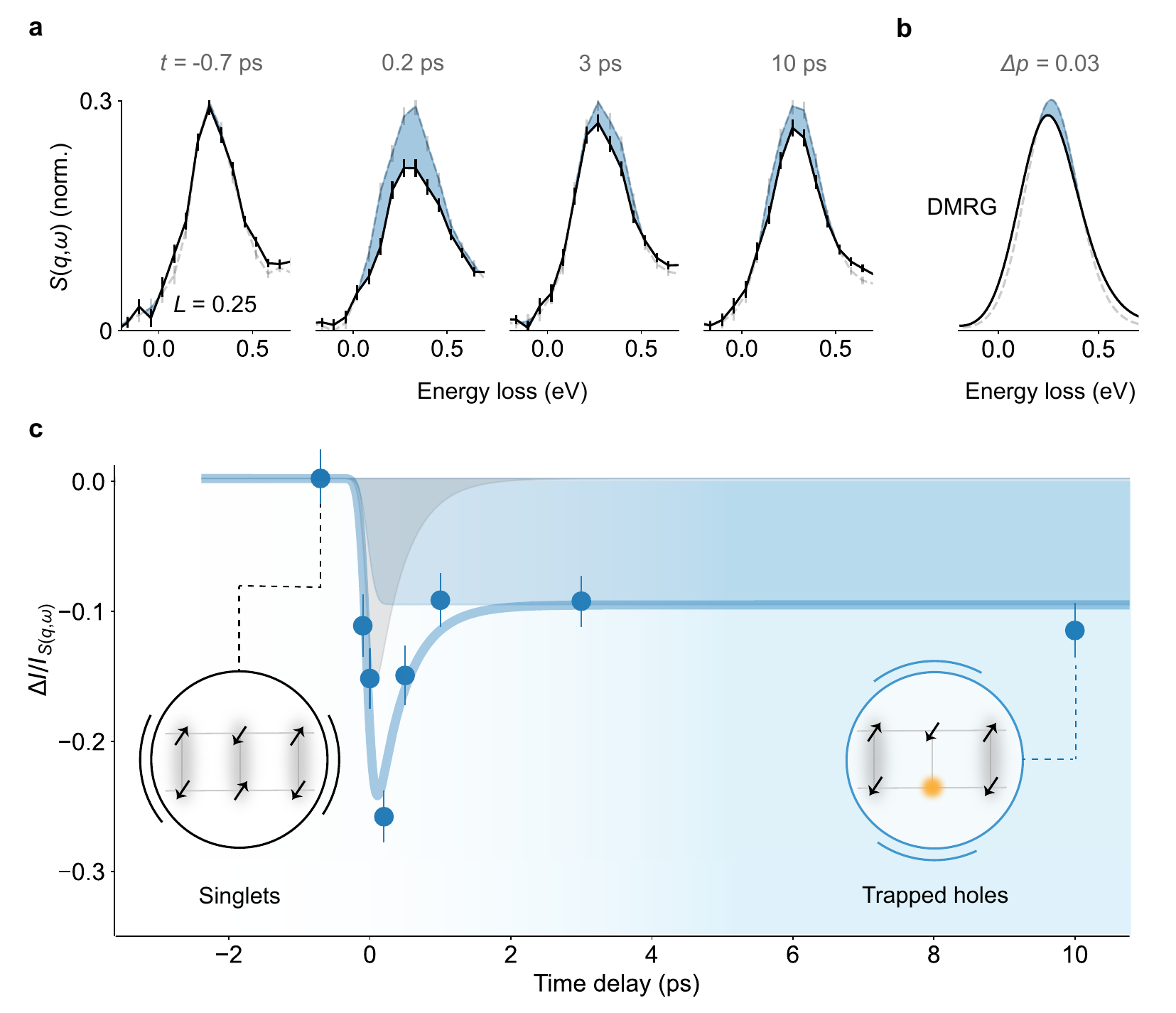}
    \caption{ {\bf Fig. ~S16. Temporal evolution of two-triplon continuum.} \textbf{a}, $S(q,\omega,t)$ at $q$ = 0.25 in (grey) and out of equilibrium (black) for selected time delays. After an initial suppression and a partial recovery, the two-triplon continuum exhibits a residual metastable intensity suppression. \textbf{b}, DMRG calculation of $S(q,\omega)$ for $p$ = 0.06 (grey, equilibrium) and $p$ = 0.09 (black, transient). The calculated two-triplon suppression matches that observed in the trRIXS spectra at long timescales. \textbf{c}, Time-dependent differential two-triplon intensity $\Delta I/I_{S(q,\omega)}$ as a function of time delay. Sketches of the chain-to-ladder hole transfer and the consequent disruption of short-range spin correlations are shown in the bubbles. Arrows denote spins, yellow circles denote holes, and the grey clouds indicate rung singlets. }
    \label{fig:trrixsDelaySnapshots}
\end{figure}

Here, we expand upon the results presented in Fig. 5 of the main text and show the temporal evolution of the two-triplon continuum in Fig. S\ref{fig:trrixsDelaySnapshots}. At longer timescales, the intensity reduction is fully captured by the metastable hole transfer, as confirmed by the agreement with DMRG calculations (Fig. S\ref{fig:trrixsDelaySnapshots}b). However, at shorter timescales, the magnetic signal exhibits a much stronger suppression, which recovers on a sub-picosecond timescale. This behavior mirrors the time evolution of the trXAS spectra (Fig. S\ref{fig:trxasDelayAndFluence}a) and is likely due to short-lived incoherent charge excitations within the ladders.

\clearpage

\section*{6. DMRG calculations}\label{sec7}

To identify spectral fingerprints of hole doping, we employ the density matrix renormalization group (DMRG) to simulate the dynamical spin structure factor $S(q, \omega)$ in a two-leg ladder system. The system length is chosen as $L=64$. Based on the quantitative experiment-theory comparisons in Ref. \cite{padma2024beyond}, this ladder system can be faithfully described by a single-band extended Hubbard model, with its Hamiltonian given by:
\begin{eqnarray}
    \mathcal{H} &=& - \sum_{jl\sigma}  t\big[c^{(l)\dagger}_{j\sigma} c^{(l)}_{j+1\sigma} + h.c.\big] -\sum_{j \sigma}  t_\perp \big[c^{(0)\dagger}_{j\sigma} c^{(1)}_{j\sigma} + h.c.\big] -\sum_{jl \sigma}  t^\prime \big[c^{(l)\dagger}_{j\sigma} c^{(1-l)}_{j+1\sigma} + h.c.\big]\nonumber\\
    &&+ U \sum_{jl} n^{(l)}_{j\uparrow} n^{(l)}_{j\downarrow} + V\sum_{j}\sum_{\sigma,\sigma^\prime}\big[ n^{(0)}_{j\sigma} n^{(1)}_{j\sigma^\prime}+\sum_{l}n^{(l)}_{j\sigma} n^{(l)}_{j+1\sigma^\prime}\big],
\end{eqnarray}
where $c^{(l)}_{j\sigma}$ ($c^{(l)\dagger}_{j\sigma}$) annihilates (creates) an electron at site $j$ on leg $l = 0,1$ with spin $\sigma = \uparrow, \downarrow$, and $n^{(l)}_{j\sigma} = c^{(l)^\dagger}_{j\sigma} c^{(l)}_{j\sigma}$ denotes the local electron density. Here, $t$ is the hopping integral between nearest neighbors along the leg, $t_\perp = 0.84t$ is the hopping along the rung, and $t^\prime=-0.3t$ is the next-nearest neighbor hopping. The on-site Hubbard repulsion is $U=8t$, and the nonlocal, likely phonon-mediated, attractive interaction is $V=-1.25t$ \cite{chen2021anomalously, wang2021phonon, padma2024beyond}. In this particular system, we find that a maximum bond dimension $D=1000$ gives truncation error magnitudes on the order of $10^{-7}$.

The dynamical spin structure factor $S(q, \omega)$ is
\begin{equation} \label{tDMRG_2leg} 
    S(q, \omega) = \int_0^{T_{\text{max}}} dt \sum_{j} \sum_{l=0,1} \left\langle G \left| \mathcal{U}(0,t)\, S^{(l)}_{j}\mathcal{U}(t,0) S^{(0)}_{j_0} \right| G \right\rangle e^{i q j}  e^{-i \omega t} e^{-t^2 / t_{\text{win}}^2}\,.
\end{equation}
Here, $S^{(l)}_j= \big[c^{(l)\dagger}_{j
\uparrow}c^{(l)}_{j\uparrow}-c^{(l)\dagger}_{j\downarrow}c^{(l)}_{j\downarrow}\big]/2$ is the spin operator at site $j$ on leg $l$, and $\mathcal{U}(t_1,t_2)$ is the time-evolution operator. To minimize boundary effects and enforce translational symmetry, the middle site $j_0=L/2$ is fixed, and the sum in Eq.~\eqref{tDMRG_2leg} runs over all site indices $j$. The two-time correlation function is evaluated using the time-dependent variational principle (TDVP), with a step size $\delta t = 0.05t^{-1}$ and a maximum time $T_{\text{max}} = 30t^{-1}$. The finite-time Fourier transform is applied using a Gaussian window function with the parameter $t_{\text{win}} = 300t^{-1}$.

\begin{figure}[t]
    \centering
    \includegraphics[width=\textwidth]{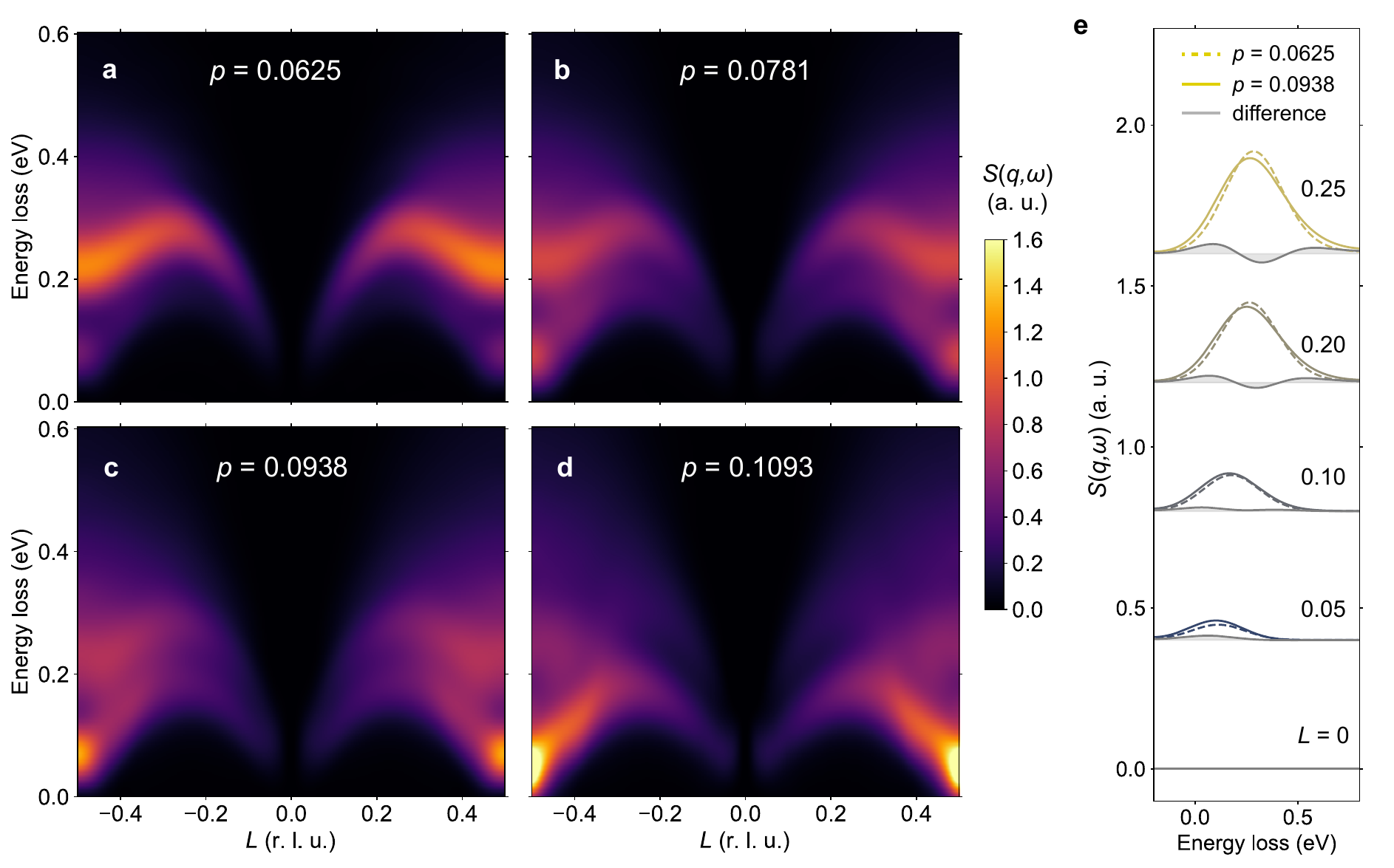}
    \caption{ {\bf Fig. ~S17. $S(q,\omega)$ calculated by DMRG.} \textbf{a-d}, $S(q,\omega)$ intensity maps for ladder hole densities, from $p$ = 0.0625/Cu$_\mathrm{L}$ to $p$ = 0.1093/Cu$_\mathrm{L}$, calculated using DMRG. \textbf{e}, $S(q,\omega)$ spectra as a function of momentum $L$, for $p$ = 0.0625/Cu$_\mathrm{L}$ and $p$ = 0.0938/Cu$_\mathrm{L}$, broadened to match the experimental energy resolution of 260 meV. The difference between the two spectra is shown in grey.}\label{SI_DMRG}
    \label{fig:s(q,w)CalculatedByDmrg}
\end{figure}

To account for the self-doping at equilibrium, a hole density $p$ = 0.0625/Cu$_\mathrm{L}$ is set as the reference, with additional hole doping and impurities introduced relative to this. As shown in Fig.~S\ref{SI_DMRG}a, the spectrum with hole density $p$ = 0.0625/Cu$_\mathrm{L}$ is most intense for the two-triplon continuum between 0.2 and 0.4\,eV. A lower excitation branch, originating from quasiparticle spin-flips, is visible between 0.05 and 0.15\,eV \cite{troyer1996properties, liu2016nature, Kumar2019ladders, Schlappa2009collective, Tseng2022crossover,padma2024beyond}. 

\subsection*{Variation of $S(q, \omega)$ with hole doping}\vspace{-3mm}

To quantify the photoinduced doping in the metastable state, we examine the $S(q, \omega)$ obtained for hole densities of $p$ = 0.0781/Cu$_\mathrm{L}$, 0.0938/Cu$_\mathrm{L}$, and 0.1093/Cu$_\mathrm{L}$ (see Fig.~S\ref{SI_DMRG}b-d). With increasing hole density, the spin spectral weight transfers from the two-triplon continuum to the quasiparticle branch. 

To enable comparison with the experimental results (Fig. 5 of the main text), we convolve the simulated spectra with a Gaussian broadening, matching the experimental energy resolution of 260 meV. We present the convolved spectra with $p$ = 0.0625/Cu$_\mathrm{L}$ and $p$ = 0.0938/Cu$_\mathrm{L}$ as a function of momentum in Fig. S\ref{SI_DMRG}e. Hole doping results in a broadening and slight red shift of the two-triplon intensity at $L > 0.10$. This manifests as a slight suppression of the peak intensity and a slight increase at the shoulders, as shown in the difference plots (filled grey area). At larger doping, such as in $p$ = 0.1093/Cu$_\mathrm{L}$, we find a significant suppression of the two-triplon continuum and transfer of spectral weight to the quasiparticle branch, which is inconsistent with the experimental observations (see Fig. S\ref{fig:trrixsT3} and Fig. 5 of the main text). Thus we find that $p$ = 0.0938/Cu$_\mathrm{L}$ offers the best agreement with the experimental $S(q, \omega)$ in the metastable state.

\begin{figure}[h]
    \centering
    \includegraphics[width=0.65\textwidth]{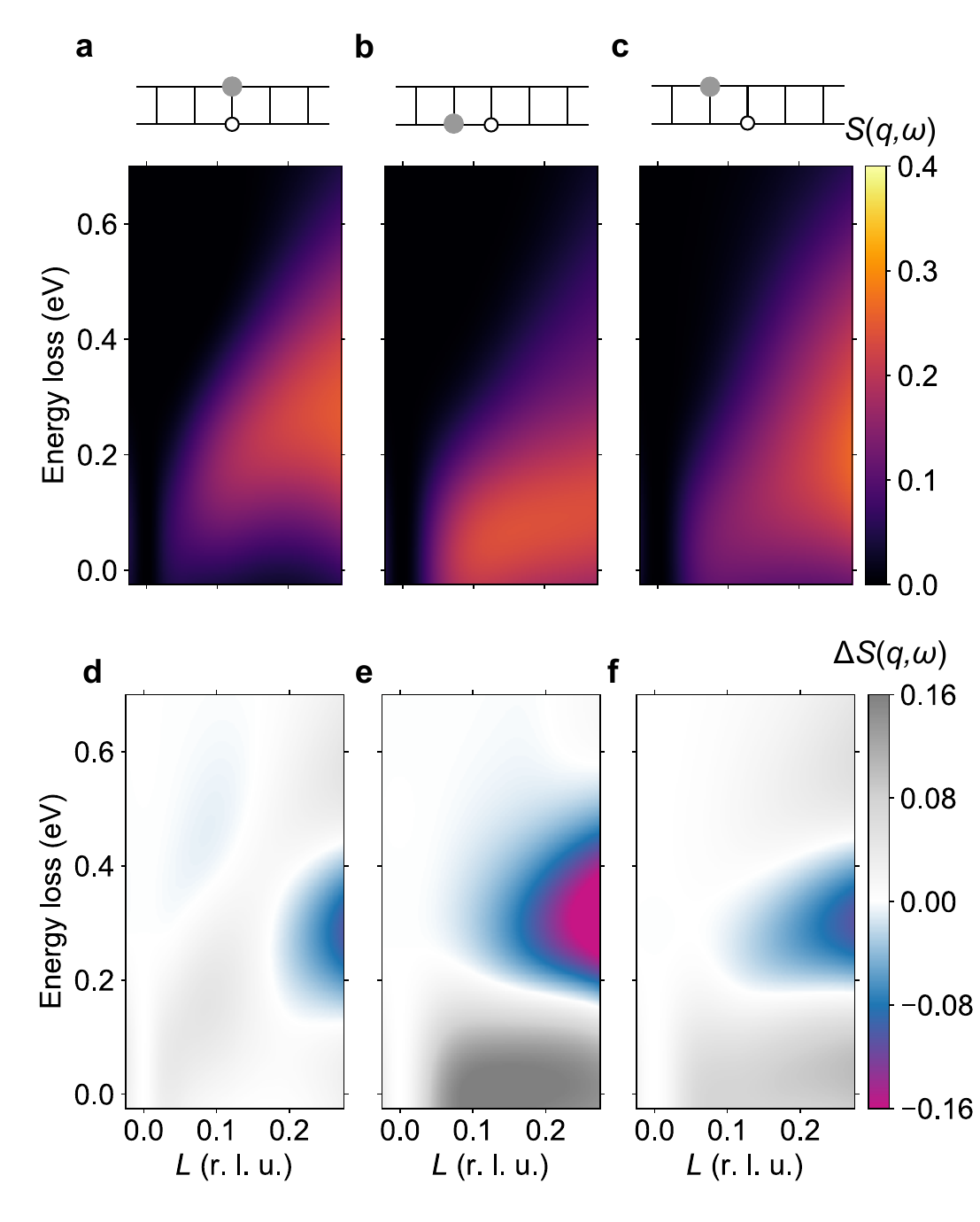} \caption{ {\bf Fig. ~S18. Spectral analysis for different localized hole configurations.} \textbf{a-c} The $S(q,\omega)$ calculated for hole density $p$ = 0.0938/Cu$_\mathrm{L}$ for three different impurity configurations.
    The upper schematics illustrate the position of the impurity site (gray dot, $(l_1,j_1)$) relative to the reference site (white dot, $(0,j_0)$) for each of the three impurity configurations. 
   \textbf{d-f} Differential intensities $\Delta S(q,\omega)$ compared to the $S(q,\omega)$ of the reference state with $p$ = 0.0625/Cu$_\mathrm{L}$ (shown in Fig.~S17\textbf{a}) for each of the three impurity configurations. }\label{SI_DMRG_impurity}
\end{figure}

\subsection*{Effect of hole localization}\vspace{-3mm}

To determine whether the observed metastable state is associated with itinerant or localized holes, we simulate $S(q,\omega)$ for $p$ = 0.0938/Cu$_\mathrm{L}$ with a local charge impurity (see Fig. 5d). To do this, we impose an additional chemical potential shift $\Delta \mu= 5t$ that is strong enough to localize the doped holes. We apply the localization potential at site $(l = l_1,j = j_1)$ near the reference point $(l=0,j=j_0)$, using the same convention as in Eq.~\eqref{tDMRG_2leg}. The modified Hamiltonian is written as
\begin{equation} \label{tDMRG_impurity}
    \mathcal{H}^\prime(l_1,j_1) = \mathcal{H} + \Delta\mu \left[n^{(l_1)}_{j_1\uparrow}+n^{(l_1)}_{j_1\downarrow}\right].
\end{equation}

We consider three inequivalent impurity locations: at sites $(1, j_0)$ (type 1), $(0, j_0 \pm 1)$ (type 2), and $(1, j_0 \pm 1)$ (type 3), as shown schematically in Fig. S\ref{SI_DMRG_impurity}. The experimental $S(q,\omega)$ in the metastable state shows a suppression centered at the two triplon peak, spanning the continuum, and without changing the shape of the dispersion (see Fig.~5d of the main text). The simulated $S(q,\omega)$ with hole localization deviates from this for all three impurity types. For impurity type 1, spectral weight is significantly suppressed at $L$ = 0.25 and transferred to lower momenta within the two triplon continuum. Impurity type 2 shows an even stronger suppression of the two triplon spectral weight, by almost a factor of two, transferred into the quasiparticle branch at lower energy loss. Impurity type 3 results in a similar redshift of spectral weight as type 2, albeit with a lower magnitude. In all cases, the suppression of the two-triplon continuum is much larger than observed in the experimental $S(q,\omega)$. Finally, to reflect the statistical average of randomly distributed impurities, we calculate a weighted average of these data based on the symmetry and equivalence of the sites. This weighted average, presented in Fig.~5d of the main text, similarly deviates from the experimental observations. Therefore our experimental results are  consistent with holes transferred into the ladder, with primarily itinerant character.

\clearpage

\section*{7. Light-driven symmetry breaking and hopping}\label{sec2}

The primary mode of interaction between the chains and ladders is via the apical oxygen atoms in the chain \cite{gotoh2003structural, deng2011structural}, as outlined in the main text. The low-energy electronic state of the ladder holes is a Zhang-Rice singlet (ZRS), composed of Cu $3d_{x^2-y^2}$ and surrounding O $2p_{x/y}$ orbitals on each plaquette, possessing approximate $D_{4h}$ symmetry. Owing to its $d$-wave character, the ZRS wavefunction switches sign under a $\pi/2$ rotation around the copper-apical oxygen axis, whereas the O $2p_z$ orbital remains invariant. As a result, the hopping integral $t_{\mathrm{ap}}$ between the ZRS and the apical oxygen, which arises from a superposition of four terms involving rotated ZRS configurations with alternating signs, cancels out. Dressing by intense in-plane electric fields breaks this symmetry by introducing an additional Peierls phase to the hopping matrix elements. This perturbation imbalances the superposition of hopping terms involving orbitals aligned with and perpendicular to the electric field, allowing for a finite $t_{\mathrm{ap}}$ between the ladder and chain. We present a detailed derivation of this result below.

\subsection*{$d$-wave symmetry of hole carriers and parity mismatch}\vspace{-3mm}

We first examine the vertical hopping between the in-plane copper $3d_{x^2-y^2}$ orbital and the approximately apical oxygen $2p_z^*$ orbital ($*$ here indicates the apical oxygen) above it. At the \textit{ab initio} level, this hopping integral is calculated as 
\begin{equation}\label{eq:tpzintegral}
    t_{dpz} = \iiint \psi_{2p_z}^*(\mathbf{r}) \,\mathcal{H}\, \psi_{3d_{x^2-y^2}}(\mathbf{r})d\mathbf{r}^3 ,
\end{equation}
where $\psi_{2p_z}(\mathbf{r})$ and $\psi_{3d_{x^2-y^2}}(\mathbf{r})$ are the Wannier wavefunctions of the apical oxygen $2p_z^*$ and in-plane copper $3d_{x^2-y^2}$ orbitals, respectively. Here, $\mathcal{H}$ is the electronic Hamiltonian and can be regarded as the single-electron part of the many-body Hamiltonian for the purpose of evaluating the hopping integral. While single-electron integrals are usually evaluated numerically in quantum chemistry, symmetry analysis can determine whether the integral will vanish or not. Since the centers of these wavefunctions are approximately aligned along the copper-apical oxygen line, we define a rotational operator $\hat{R}(\theta)$ about this axis. The $3d_{x^2-y^2}$ and $2p_z$ orbitals exhibit different parities under $\pi/2$ rotations
\begin{align}
    \hat{R}\left(\frac{\pi}{2} \right) \psi_{2p_z}(\mathbf{r}) = \psi_{2p_z}(\mathbf{r})\,,  \quad\textrm{and}\quad \hat{R}\left(\frac{\pi}{2} \right) \psi_{3d_{x^2-y^2}}(\mathbf{r}) = -\psi_{3d_{x^2-y^2}}(\mathbf{r}) .
\end{align}
If we assume a perfect $D_{4h}$ symmetry, the Hamiltonian will be symmetric under these rotations, namely $\hat{R}\left(\frac{\pi}{2} \right) \mathcal{H} = \mathcal{H}$. Under such symmetry, we can then split the hopping integral as shown below.
\begin{align}
    t_{dpz} = & \quad \frac{1}{4} \Big[\iiint \psi_{2p_z}^*(\mathbf{r}) \mathcal{H} \psi_{3d_{x^2-y^2}}(\mathbf{r})d\mathbf{r}^3 + \hat{R}\left(\frac{\pi}{2} \right)\iiint \psi_{2p_z}^*(\mathbf{r}) \,\mathcal{H}\, \psi_{3d_{x^2-y^2}}(\mathbf{r})d\mathbf{r}^3 \nonumber\\
    &+ \hat{R}\left(\pi \right)\iiint \psi_{2p_z}^*(\mathbf{r}) \,\mathcal{H}\, \psi_{3d_{x^2-y^2}}(\mathbf{r})d\mathbf{r}^3 + \hat{R}\left(\frac{3\pi}{2} \right)\iiint \psi_{2p_z}^*(\mathbf{r}) \,\mathcal{H}\, \psi_{3d_{x^2-y^2}}(\mathbf{r})d\mathbf{r}^3
    \Big] \nonumber\\
    = & \quad\frac{1}{4} \Big[\iiint \psi_{2p_z}^*(\mathbf{r}) \,\mathcal{H}\, \psi_{3d_{x^2-y^2}}(\mathbf{r})d\mathbf{r}^3 -\iiint \psi_{2p_z}^*(\mathbf{r}) \,\mathcal{H}\, \psi_{3d_{x^2-y^2}}(\mathbf{r})d\mathbf{r}^3 \nonumber\\
    &+ \iiint \psi_{2p_z}^*(\mathbf{r}) \,\mathcal{H}\, \psi_{3d_{x^2-y^2}}(\mathbf{r})d\mathbf{r}^3 -\iiint \psi_{2p_z}^*(\mathbf{r}) \,\mathcal{H}\, \psi_{3d_{x^2-y^2}}(\mathbf{r})d\mathbf{r}^3
    \Big] \nonumber\\
    = & \quad 0\,.
\end{align}
Owing to the perfect cancellation of the four possible hopping integrals, the apical oxygen $2p_z^*$ orbital and the in-plane copper $3d_{x^2-y^2}$ orbital are completely decoupled from each other.

While our ladder system does not perfectly satisfy the $D_{4h}$ symmetry due to half-lattice shifts between neighboring ladders, the four oxygen atoms surrounding each copper atom still form an approximately square plaquette, with the Cu-O bond length varying within the range 1.96 - 2.00\,\AA\ along the leg direction and 1.92 - 1.97\,\AA\ along the rung direction. Moreover, the single-electron integrals between the valence orbitals of adjacent ladders are zero due to the parity mismatch between the orbitals, which effectively suppresses interladder coupling. Consequently, although the hopping $t_{dpz}$ is not exactly zero in Sr$_{14-x}$Ca$_{x}$Cu$_{24}$O$_{41}$, it is small compared to that induced by the laser (discussed further in the next subsection).

Since cuprates are charge transfer compounds, with the in-plane Cu-O states exhibiting covalent character, low-energy electronic states due to hole doping are distributed across the Cu and in-plane O atoms\,\cite{zaanen1989charged, imada1998metal}. This implies that we also need to consider hopping involving O $2p$ orbitals along the $x$ (leg) and $y$ (rung) directions. Unlike the central Cu $3d$ orbitals, these O orbitals are not aligned with the apical oxygen, and exhibit finite hopping integrals denoted as $t_{pp}$. Our single-unit-cell \textit{ab initio} calculation indicates a hopping $t_{pp}\sim 0.27$\,eV, consistent with the value in high-$T_\mathrm{C}$ cuprates such as La$_{2-x}$Ba$_{x}$CuO$_{4}$ \cite{mcmahan1990cuprate, mcmahan1988calculated}. Hence, in self-doped Sr$_{14-x}$Ca$_{x}$Cu$_{24}$O$_{41}$ ladders, in-plane and apical oxygens can, in principle, allow chain-to-ladder hopping. 

However, doped carriers in the O orbitals form Zhang-Rice singlets that obey the same symmetry as the central $3d_{x^2-y^2}$ orbital \cite{zhang1988effective}. The other bonding and antibonding states are at higher energies and remain unoccupied \cite{kung2016characterizing}. To estimate the effective hopping between the Zhang-Rice singlet and the apical oxygen, we define an effective oxygen wavefunction on each copper site as a symmetric superposition
\begin{equation}
    \psi_p(\mathbf{r};\mathbf{R}) = \frac{1}{2} \Big[\psi_{p_x}(\mathbf{r};\mathbf{R}-\frac{a_0}{2}\hat{x})  - \psi_{p_x}(\mathbf{r};\mathbf{R}-\frac{a_0}{2}\hat{x}) + \psi_{p_y}(\mathbf{r};\mathbf{R}-\frac{a_0}{2}\hat{y}) - \psi_{p_y}(\mathbf{r};\mathbf{R}+\frac{a_0}{2}\hat{y})\Big].
\end{equation}
This effective wavefunction is antisymmetric under rotations of $\pi/2$, as
\begin{equation}
    \hat{R}\left( \frac{\pi}{2}\right)\psi_p(\mathbf{r};\mathbf{R}) =-\psi_p(\mathbf{r};\mathbf{R}),
\end{equation}
which is the same symmetry observed in the copper $3d_{x^2-y^2}$ orbital. As a consequence, the hopping between the in-ladder oxygen $p_{x/y}$ orbital and the apical oxygen ${p_z^*}$ orbital also vanishes:
\begin{align}
    t_{ppz} = & \quad\frac{1}{4} \Big[\iiint \psi_{2p_z}^*(\mathbf{r}) \,\mathcal{H}\, \psi_{p}(\mathbf{r})d\mathbf{r}^3 -\iiint \psi_{2p_z}^*(\mathbf{r}) \,\mathcal{H}\, \psi_p(\mathbf{r})d\mathbf{r}^3 \nonumber\\
    &+ \iiint \psi_{2p_z}^*(\mathbf{r}) \,\mathcal{H}\, \psi_p(\mathbf{r})d\mathbf{r}^3 -\iiint \psi_{2p_z}^*(\mathbf{r}) \,\mathcal{H}\, \psi_p(\mathbf{r})d\mathbf{r}^3
    \Big] \nonumber\\
    = & \quad 0\,.
\end{align}

Since both $t_{dpz}$ and $t_{pp}$ vanish, the vertical charge transfer between the ladder and the apical oxygen also vanishes. 

\subsection*{Chain-to-ladder hopping triggered by laser field}\vspace{-3mm}
The $D_{4h}$ symmetry may be broken by the pump electric field, leading to a light-induced charge transfer between chains and ladders. While the symmetry of the Hamiltonian is preserved when we apply an out-of-plane polarized ($E \parallel b$) pump to the system, the $C_4$ symmetry of the CuO$_4$ plaquette can be broken by an in-plane polarized ($E \parallel a$ or $E \parallel c$) electric field. In the weak field limit, we linearize the field dependence of the Hamiltonian as
\begin{equation}
    \mathcal{H}(\mathbf{A}) = \mathcal{H}(\mathbf{A}=0) -\frac{e}{2m_ec}\sum_j\left[i\hbar\nabla_j\cdot\mathbf{A}(\mathbf{r_j}) + h.c.\right] + O(\mathbf{A}^2),
\end{equation}
and, choosing the polarization as $\mathbf{A}(\mathbf{r_j}) = A_0(\mathbf{r})\hat{x}$, we rewrite the hopping as 
\begin{equation}\label{eq:hoppingIntegral}
    t_{ppz} =  \iiint \psi_{2p_z}^*(\mathbf{r}) \left[ \mathcal{H}(A_0=0) -\frac{i\hbar e}{m_ec} [A_0(\mathbf{r})\frac{\partial}{\partial x} + \frac{\partial A_0(\mathbf{r})}{\partial x}] + O(A_0^2) \right] \psi_{p}(\mathbf{r})d\mathbf{r}^3.
\end{equation}
As discussed in the prior section, the first equilibrium term vanishes. Since the linear term exhibits odd parity, and both $\psi_{2p_z}$ and $\psi_{p}$ have even parity, this term will also vanish, implying that the hopping in the presence of an optical pump is proportional to the square of the field strength. This conclusion also applies to $t_{dpz}$, the hopping integral between the Cu $3d_{x^2-y^2}$ orbital and the apical O $2p_z$ orbital.

The second-order field dependence for the oxygen orbitals can be evaluated by using the Peierls substitution due to the separation of charge centers. The effect of the external vector potential is simplified into a location-dependent phase factor in the hopping integral. For two Bloch wavefunctions centered at sites $\mathbf{r}_i$ and $\mathbf{r}_j$, the hopping integral becomes
\begin{equation}
    t(\mathbf{r}_i,\mathbf{r}_j) \rightarrow t(\mathbf{r}_i,\mathbf{r}_j) \text{exp}\left( \frac{ie}{\hbar c}\int_{\mathbf{r}_i}^{\mathbf{r}_j}\mathbf{A}(\mathbf{r})\cdot d\mathbf{r} \right).
\end{equation}
Thus, we estimate the light-driven hopping integral between the bonding oxygen wavefunctions in the ladder and the $2p_z$ orbital of the apical oxygen in the chain as
\begin{align}
    t_{ppz} =  \frac{1}{2}\big(t_{pp} + t_{pp} - t_{pp}e^{-i\frac{eA_0a_0}{2\hbar c}} - t_{pp}e^{i\frac{eA_0a_0}{2\hbar c}}\big) =  t_{pp} \left[1-\cos\left(\frac{eA_0a_0}{2\hbar c}\right)\right] \simeq  \frac{1}{2}\left(\frac{eA_0a_0}{2\hbar c}\right)^2t_{pp} \propto A_0^2.
\end{align}
This result shows that the $t_{ppz}$ matrix element, which vanishes at equilibrium, becomes nonzero due to the applied electric field, and its amplitude scales linearly with the field intensity. Our \textit{ab initio} simulation using a CuO$_6^{9-}$ cluster, that is a CuO$_6^{10-}$ ionic compound with a doped hole, confirms this square scaling with the electric field, giving $t_{ppz}\approx1.08\,$meV for for $E=7$\,MV/cm.

One may further consider the steady-state renormalization of the hopping matrix elements in the presence of the optical pump. The effective period-averaged hopping integral can be expressed as
\begin{align}
    t_{ppz}^{\rm(eff)} = \frac{2\pi}{\Omega}\int_0^{2\pi/\Omega}  t_{pp} \left[1-\cos\left(\frac{ea_0A_0\cos(\Omega \tau)}{2\hbar c}\right) \right]\,d\tau = \left[1-\mathcal{J}_0\left(\frac{ea_0A_0}{2\hbar c}\right)\right]t_{pp}\,.
\end{align}
Here, $\mathcal{J}_0(z)$ is the Bessel function of the first kind. In the weak-pump limit, the hopping integral is also proportional to $A_0^2$ at the leading order. Incorporating this value into the \textit{ab initio} simulation, we obtain $t_{ppz}^{\rm(eff)}\approx 0.54\,$meV for for $E=7$\,MV/cm.

\subsection*{Pump-induced energy shift of ladder and chain orbitals}\vspace{-3mm}

The optically-dressed hopping determines the rate of the hole transfer and establishes metastability. On the other hand, the overall number of transferred carriers is determined by the difference in the optically-dressed orbital energies.

Similar to the one-electron integral Eq.~\eqref{eq:hoppingIntegral}, the energy of any of the relevant orbitals in the presence of the vector potential is given by
\begin{eqnarray}\label{eq:orbitalEnergy}
    \epsilon_{\alpha}(\mathbf{A}) &=& \iiint \psi_{\alpha}(\mathbf{r}) \bigg[\hat H(\mathbf{A} = 0) -i \frac{e\hbar}{2m_e c}\frac{\partial A_0(\mathbf{r})}{\partial x}-iA_0(\mathbf{r}) \frac{e\hbar}{m_e c}\frac{\partial}{\partial x}  + O(\mathbf{A}^2) \bigg] \psi_{\alpha}(\mathbf{r}) d\mathbf{r}^3\,,
\end{eqnarray}
where $\alpha=3d_{x^2-y^2}$, $2p_x$, $2p_y$, or $2p_z^*$. The first term of the integral gives the equilibrium orbital energy, denoted as $\epsilon_{\alpha}^{\rm(eq)}$. Since all relevant orbital wavefunctions are either even or odd under inversion, their squares (density) are always even, leading to the vanishing of the second and third terms. Therefore, all orbital energies, and hence the total hole transfer in the metastable state, are expected to change quadratically with the external field $\mathbf{A}$, or equivalently, the electric field strength $\mathbf{E}=-\partial \mathbf{A}/\partial t$.

\begin{figure}
    \centering
    \includegraphics[width=0.6\linewidth]{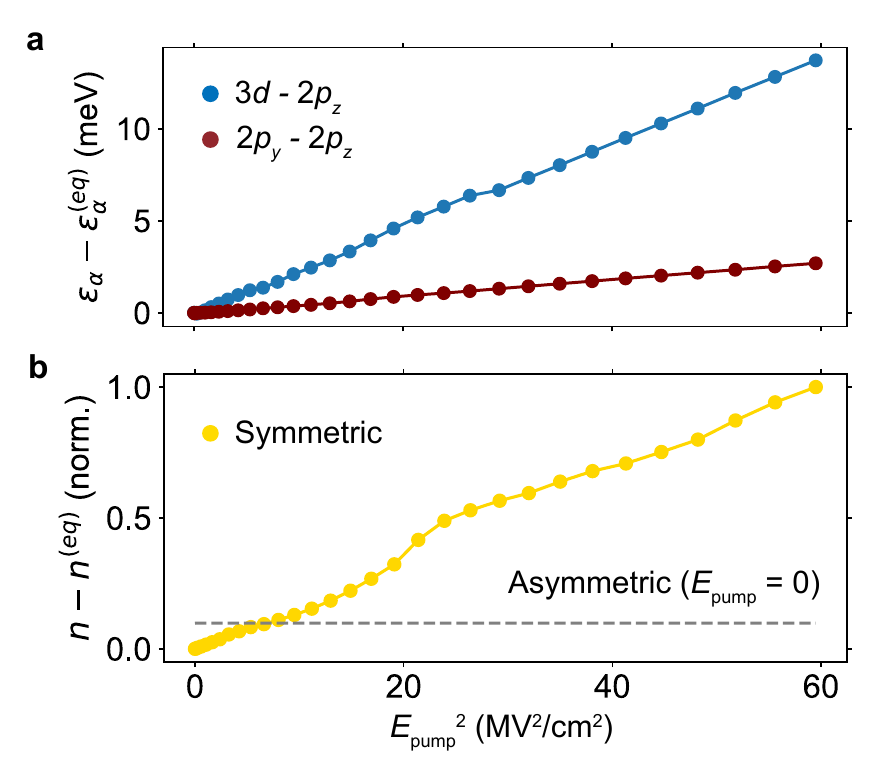}
    \caption{{\bf Fig. ~S19. Quadratic dependence of site energy and charge density with the external field.} \textbf{a}, Energy difference between $3d_{x^2-y^2}$ and $2p_z$ orbitals (blue), and between $2p_y$ and $2p_z$ orbitals (maroon) as function of the squared electric field for a CuO$_6^{9-}$ cluster with a symmetrical plaquette (Cu-O bond length = $1.96$\AA). \textbf{b}, Normalized charge density change on the apical O atom as function of the squared electric field. The grey dashed line denotes the normalized charge density for an unpumped asymmetric plaquette with Cu-O bond length of $1.945$\AA\ along the rung and $1.98$\AA\ along the leg.}
    \label{fig:siteEnergyDeltaPartialCharge}
\end{figure}

We verify this quadratic dependence by simulating the Hartree-Fock orbital energy of a CuO$_6^{9-}$ cluster. As shown in Fig.~\ref{fig:siteEnergyDeltaPartialCharge}a, the energy difference between the Cu $3d_{x^2-y^2}$ and apical oxygen $2p_z$ orbitals increases linearly with $|\mathbf{E}|^2$, as does the difference between the in-plane and apical oxygen orbitals. In both cases, the orbital energy of the apical oxygen is lowered in the presence of the laser field, leading to a charge transfer from the ladder to the chain. We further validate this conclusion by simulating the partial charge of the apical oxygens in the CuO$_6^{9-}$ cluster with a correlation-consistent basis (here, we used the cc-pVTZ basis). While these single-unit-cell simulations cannot model itinerant behavior, which is necessary for a quantitative description of the observed charge transfer, they can reliably capture qualitative trends. The light-induced electron density increases quadratically with the pump field, as shown in Fig.~\ref{fig:siteEnergyDeltaPartialCharge}b. This is consistent with the experimental pump field dependence of the differential ladder and chain XAS intensities in the metastable state. 

To provide an order-of-magnitude estimation of the total charge transfer in the experimental system with itinerant electrons, we evaluate the compressibility of the ladder system by ignoring interactions. We find that the total light-induced energy shift obtained from our ab initio calculations results in a 2\% charge transfer, consistent with the experimentally observed 3\% light-induced doping, further corroborating the mechanism of hole transfer and metastability.

\subsection*{Hole transfer due to deviations from the D$_{4h}$ symmetry}\vspace{-3mm}

Our theoretical picture assumes perfect $D_{4h}$ symmetry at equilibrium for simplicity. However, the quasi-1D crystal structure of Sr$_{14-x}$Ca$_{x}$Cu$_{24}$O$_{41}$ slightly breaks this symmetry at equilibrium. Specifically, the Cu-O bond length along the leg direction is slightly larger than that on the rung. To estimate the impact of this asymmetry, we calculate the charge transfer resulting from a breaking of the D$_{4h}$ symmetry at equilibrium. We take the average bond length to be $\sim1.98\:$\AA~ along the leg direction and $\sim1.945\:$\AA~ along the rungs. Since the geometric asymmetry may lead to different configurations of molecular orbitals, we do not compare the orbital energy and instead directly focus on its effect on the chain-to-ladder hole transfer. As shown in Fig.~\ref{fig:siteEnergyDeltaPartialCharge}b, this slight geometric asymmetry contributes an initial charge transfer that is an order of magnitude smaller than that due to the pump field. Therefore, we neglect equilibrium deviations from the $D_{4h}$ symmetry in the discussion of our results.

\clearpage


\bibliography{sn-bibliography}